%% file: main.tex
\pdfoutput=1
\documentclass[12pt,a4paper]{article}
\input{bhhh-symbols-def}
\usepackage{ifthen} 
\newboolean{pdflatex}
\setboolean{pdflatex}{true} 

\newboolean{articletitles}
\setboolean{articletitles}{true} 

\newboolean{uprightparticles}
\setboolean{uprightparticles}{false} 

\def\paperauthors{LHCb collaboration} 
\def\paperasciititle{Amplitude analysis of the D+ -> pi- pi+ pi+  decay and measurement of the pi-pi+ S-wave amplitude} 
\def\papertitle{Amplitude analysis of the $D^+\to\pi^- \pi^+\pi^+$ decay and measurement of the $\pi^-\pi^+$ S-wave amplitude} 
\def\paperkeywords{{High Energy Physics}, {LHCb}} 
\def\papercopyright{\the\year\ CERN for the benefit of the LHCb collaboration} 
\def\paperlicence{CC BY 4.0 licence}
\def\paperlicenceurl{https://creativecommons.org/licenses/by/4.0/}

\input{preamble}
\usepackage{longtable} 
\usepackage{booktabs} 
\usepackage{multirow}

\begin{document}

\def\Dppp         {\decay{\Dp}{\pim\pip\pip}}
 \def\Dsppp        {\decay{\Ds}{\pim\pip\pip}}
 \def\DKKp         {\decay{\Dp}{\Km\Kp\pip}}
\def\DsKKp        {\decay{\Ds}{\Km\Kp\pip}}
\def\ppp           {\pim\pip\pip}

\renewcommand{\thefootnote}{\fnsymbol{footnote}}
\setcounter{footnote}{1}

\input{title-LHCb-PAPER}


\renewcommand{\thefootnote}{\arabic{footnote}}
\setcounter{footnote}{0}

\cleardoublepage


\pagestyle{plain} 
\setcounter{page}{1}
\pagenumbering{arabic}



\input{body}

\input{acknowledgements}




\addcontentsline{toc}{section}{References}
\bibliographystyle{LHCb}
\bibliography{main,standard,LHCb-PAPER,LHCb-CONF,LHCb-DP,LHCb-TDR}

\newpage
\input{Authorship_LHCb-PAPER-2022-016}

\end{document}

%% file: preamble.tex
\usepackage[top=1in, bottom=1.25in, left=1in, right=1in]{geometry}

\usepackage{microtype}
\usepackage{lineno}  
\usepackage{xspace} 
\usepackage{caption} 

\usepackage{graphicx}  
\usepackage{color}
\usepackage{colortbl}
\graphicspath{{./figs/}} 

\usepackage{amsmath} 
\usepackage{amssymb}
\usepackage{amsfonts}
\usepackage{upgreek} 

\newcommand*\patchAmsMathEnvironmentForLineno[1]{%
\expandafter\let\csname old#1\expandafter\endcsname\csname #1\endcsname
\expandafter\let\csname oldend#1\expandafter\endcsname\csname
end#1\endcsname
 \renewenvironment{#1}%
   {\linenomath\csname old#1\endcsname}%
   {\csname oldend#1\endcsname\endlinenomath}%
}
\newcommand*\patchBothAmsMathEnvironmentsForLineno[1]{%
  \patchAmsMathEnvironmentForLineno{#1}%
  \patchAmsMathEnvironmentForLineno{#1*}%
}
\AtBeginDocument{%
\patchBothAmsMathEnvironmentsForLineno{equation}%
\patchBothAmsMathEnvironmentsForLineno{align}%
\patchBothAmsMathEnvironmentsForLineno{flalign}%
\patchBothAmsMathEnvironmentsForLineno{alignat}%
\patchBothAmsMathEnvironmentsForLineno{gather}%
\patchBothAmsMathEnvironmentsForLineno{multline}%
\patchBothAmsMathEnvironmentsForLineno{eqnarray}%
}

\usepackage{hyperxmp}

\usepackage[pdftex,
            pdfauthor={\paperauthors},
            pdftitle={\paperasciititle},
            pdfkeywords={\paperkeywords},
            pdfcopyright={Copyright (C) \papercopyright},
            pdflicenseurl={\paperlicenceurl}]{hyperref}

\usepackage[colorinlistoftodos,textsize=scriptsize]{todonotes}

\usepackage[bottom,flushmargin,hang,multiple]{footmisc}

\usepackage[all]{hypcap} 

\input{lhcb-symbols-def} 

\usepackage{cite} 
\usepackage{mciteplus}

%% file: lhcb-symbols-def.tex
\usepackage{xspace} 
\usepackage{upgreek}

\def\lhcb   {\mbox{LHCb}\xspace}

\def\MagUp {\mbox{\em Mag\kern -0.05em Up}\xspace}

\ifthenelse{\boolean{uprightparticles}}%
{

 \def\Ppi         {\ensuremath{\uppi}\xspace}

 \def\PDelta      {\ensuremath{\Delta}\xspace}                 
 \def\PXi         {\ensuremath{\Xi}\xspace}                 
 \def\PLambda     {\ensuremath{\Lambda}\xspace}                 
 \def\PSigma      {\ensuremath{\Sigma}\xspace}                 
 \def\POmega      {\ensuremath{\Omega}\xspace}                 
 \def\PUpsilon    {\ensuremath{\Upsilon}\xspace}
 \let\oldPi\Pi
 \def\PPi         {\ensuremath{\oldPi}\xspace}

 \def\PB      {\ensuremath{\mathrm{B}}\xspace}                 
                  
 \def\PD      {\ensuremath{\mathrm{D}}\xspace}

 \def\PK      {\ensuremath{\mathrm{K}}\xspace}

 \def\PW      {\ensuremath{\mathrm{W}}\xspace}

 \def\Pe      {\ensuremath{\mathrm{e}}\xspace}

 \def\Pi      {\ensuremath{\mathrm{i}}\xspace}

 \def\Ps      {\ensuremath{\mathrm{s}}\xspace}

 \def\thebaroffset{0.0em}
}
{

 \def\Ppi         {\ensuremath{\pi}\xspace}

 \mathchardef\PDelta="7101
 \mathchardef\PXi="7104
 \mathchardef\PLambda="7103
 \mathchardef\PSigma="7106
 \mathchardef\POmega="710A
 \mathchardef\PUpsilon="7107
 \mathchardef\PPi="7105
                  
 \def\PB      {\ensuremath{B}\xspace}                 
                  
 \def\PD      {\ensuremath{D}\xspace}

 \def\PK      {\ensuremath{K}\xspace}

 \def\PW      {\ensuremath{W}\xspace}

 \def\Pe      {\ensuremath{e}\xspace}

 \def\Pi      {\ensuremath{i}\xspace}

 \def\Ps      {\ensuremath{s}\xspace}

 \def\thebaroffset{0.18em}
}
\newcommand{\offsetoverline}[2][\thebaroffset]{\kern #1\overline{\kern -#1 #2}}%

\makeatletter
\ifcase \@ptsize \relax
  \newcommand{\miniscule}{\@setfontsize\miniscule{4}{5}}
\or
  \newcommand{\miniscule}{\@setfontsize\miniscule{5}{6}}
\or
  \newcommand{\miniscule}{\@setfontsize\miniscule{5}{6}}
\fi
\makeatother

\DeclareRobustCommand{\optbar}[1]{\shortstack{{\miniscule (\rule[.5ex]{1.25em}{.18mm})}
  \\ [-.7ex] $#1$}}

\def\ep         {{\ensuremath{\Pe^+}}\xspace}

\def\W      {{\ensuremath{\PW}}\xspace}

\def\squark    {{\ensuremath{\Ps}}\xspace}

\def\pion   {{\ensuremath{\Ppi}}\xspace}

\def\pip    {{\ensuremath{\pion^+}}\xspace}
\def\pim    {{\ensuremath{\pion^-}}\xspace}

\def\kaon    {{\ensuremath{\PK}}\xspace}
\def\Kbar    {{\ensuremath{\offsetoverline{\PK}}}\xspace}
\def\Kb      {{\ensuremath{\Kbar}}\xspace}
\def\KorKbar {\kern \thebaroffset\optbar{\kern -\thebaroffset \PK}{}\xspace}

\def\Kp      {{\ensuremath{\kaon^+}}\xspace}
\def\Km      {{\ensuremath{\kaon^-}}\xspace}

\def\KS      {{\ensuremath{\kaon^0_{\mathrm{S}}}}\xspace}

\def\D       {{\ensuremath{\PD}}\xspace}

\def\DorDbar {\kern \thebaroffset\optbar{\kern -\thebaroffset \PD}\xspace}

\def\Dp      {{\ensuremath{\D^+}}\xspace}
\def\Dm      {{\ensuremath{\D^-}}\xspace}

\def\DpDm    {\ensuremath{\Dp {\kern -0.16em \Dm}}\xspace}

\def\Ds      {{\ensuremath{\D^+_\squark}}\xspace}

\def\B       {{\ensuremath{\PB}}\xspace}

\def\BorBbar {\kern \thebaroffset\optbar{\kern -\thebaroffset \PB}\xspace}

\def\Bd      {{\ensuremath{\B^0}}\xspace}

\def\BdorBdbar {\kern \thebaroffset\optbar{\kern -\thebaroffset \Bd}\xspace}

\def\Bs      {{\ensuremath{\B^0_\squark}}\xspace}

\def\BsorBsbar {\kern \thebaroffset\optbar{\kern -\thebaroffset \Bs}\xspace}

\def\Y#1S{\ensuremath{\PUpsilon{(#1S)}}\xspace}

\def\LorLbar     {\kern \thebaroffset\optbar{\kern -\thebaroffset \PLambda}\xspace}

\newcommand{\decay}[2]{\ensuremath{#1\!\to #2}\xspace} 

\def\to                 {\ensuremath{\rightarrow}\xspace}

\def\AT#1     {\ensuremath{A_{\mathrm{T}}^{#1}}\xspace}           

\def\C#1      {\ensuremath{\mathcal{C}_{#1}}\xspace}                       
\def\Cp#1     {\ensuremath{\mathcal{C}_{#1}^{'}}\xspace}                    
\def\Ceff#1   {\ensuremath{\mathcal{C}_{#1}^{\mathrm{(eff)}}}\xspace}        
\def\Cpeff#1  {\ensuremath{\mathcal{C}_{#1}^{'\mathrm{(eff)}}}\xspace}       
\def\Ope#1    {\ensuremath{\mathcal{O}_{#1}}\xspace}                       
\def\Opep#1   {\ensuremath{\mathcal{O}_{#1}^{'}}\xspace}                    

\newcommand{\nospaceunit}[1]{\ensuremath{\text{#1}}}       
\newcommand{\aunit}[1]{\ensuremath{\text{\,#1}}}       

\newcommand{\tev}{\aunit{Te\kern -0.1em V}\xspace}
\newcommand{\gev}{\aunit{Ge\kern -0.1em V}\xspace}
\newcommand{\mev}{\aunit{Me\kern -0.1em V}\xspace}
\newcommand{\kev}{\aunit{ke\kern -0.1em V}\xspace}
\newcommand{\ev}{\aunit{e\kern -0.1em V}\xspace}
 
\newcommand{\mevc}{\ensuremath{\aunit{Me\kern -0.1em V\!/}c}\xspace}
\newcommand{\gevc}{\ensuremath{\aunit{Ge\kern -0.1em V\!/}c}\xspace}
\newcommand{\mevcc}{\ensuremath{\aunit{Me\kern -0.1em V\!/}c^2}\xspace}
\newcommand{\gevcc}{\ensuremath{\aunit{Ge\kern -0.1em V\!/}c^2}\xspace}

\def\mum  {\ensuremath{\,\upmu\nospaceunit{m}}\xspace}

\def\fb   {\ensuremath{\aunit{fb}}\xspace}
\def\invfb   {\ensuremath{\fb^{-1}}\xspace}

\newcommand{\chisq}{\ensuremath{\chi^2}\xspace}

\newcommand{\chisqip}{\ensuremath{\chi^2_{\text{IP}}}\xspace}

\def\gsim{{~\raise.15em\hbox{$>$}\kern-.85em
          \lower.35em\hbox{$\sim$}~}\xspace}
\def\lsim{{~\raise.15em\hbox{$<$}\kern-.85em
          \lower.35em\hbox{$\sim$}~}\xspace}

\def\pt         {\ensuremath{p_{\mathrm{T}}}\xspace}

\def\evtgen     {\mbox{\textsc{EvtGen}}\xspace}

\def\geant      {\mbox{\textsc{Geant4}}\xspace}

\def\photos     {\mbox{\textsc{Photos}}\xspace}

\def\pythia     {\mbox{\textsc{Pythia}}\xspace}

\def\tell1  {TELL1\xspace}
\def\ukl1   {UKL1\xspace}

\newcommand{\lhcborcid}[1]{\href{https://orcid.org/#1}{\hspace*{0.1em}\raisebox{-0.45ex}{\includegraphics[width=1em]{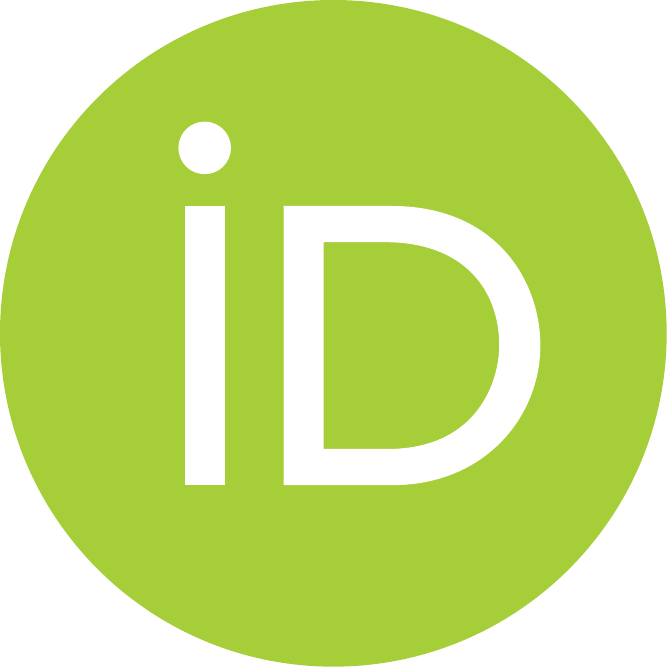}}}}


%% file: title-LHCb-PAPER.tex

\begin{titlepage}
\pagenumbering{roman}

\vspace*{-1.5cm}
\centerline{\large EUROPEAN ORGANIZATION FOR NUCLEAR RESEARCH (CERN)}
\vspace*{1.5cm}
\noindent
\begin{tabular*}{\linewidth}{lc@{\extracolsep{\fill}}r@{\extracolsep{0pt}}}
\ifthenelse{\boolean{pdflatex}}
{\vspace*{-1.5cm}\mbox{\!\!\!\includegraphics[width=.14\textwidth]{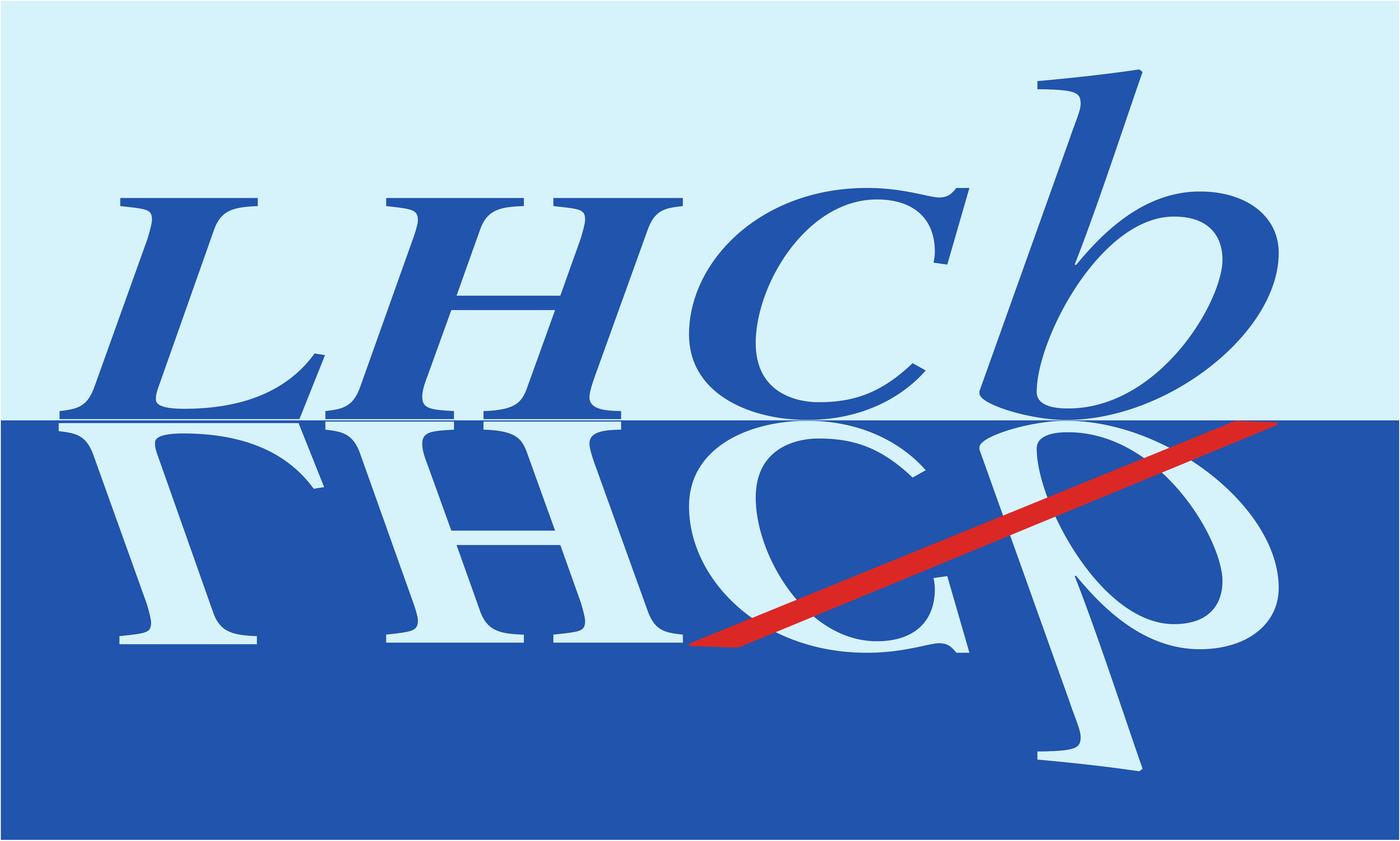}} & &}%
{\vspace*{-1.2cm}\mbox{\!\!\!\includegraphics[width=.12\textwidth]{figs/lhcb-logo.eps}} & &}%
\\
 & & CERN-EP-2022-136 \\  
 & & LHCb-PAPER-2022-016 \\  
 & & \today \\ 
 & & \\
\end{tabular*}

\vspace*{4.0cm}

{\normalfont\bfseries\boldmath\huge
\begin{center}
  \papertitle 
\end{center}
}

\vspace*{2.0cm}

\begin{center}
\paperauthors\footnote{Authors are listed at the end of this paper.}
\end{center}

\vspace{\fill}

\begin{abstract}
  \noindent
An amplitude analysis of the $D^+ \to \pi^- \pi^+ \pi^+$ decay is performed with a sample corresponding to 1.5\invfb of integrated luminosity of $pp$ collisions at a centre-of-mass energy $\sqrt{s}=8$\tev collected by the LHCb detector in 2012. The sample contains approximately six hundred thousand candidates with a signal purity of $95\%$. The resonant structure is studied through a fit to the Dalitz plot where the $\pi^- \pi^+$ S-wave amplitude is extracted as a function of $\pim\pip$ mass, and spin-1 and spin-2 resonances are included coherently  through an isobar model. The S-wave component is found to be dominant, followed by the $\rho(770)^0\pip$ and $f_2(1270)\pip$ components. A small contribution from the $\omega(782)\to\pim\pip$ decay is seen for the first time in the \Dppp decay.

\end{abstract}

\vspace*{2.0cm}

\begin{center}
  Published in JHEP 06 (2023) 044
\end{center}

\vspace{\fill}

{\footnotesize 
\centerline{\copyright~\papercopyright. \href{\paperlicenceurl}{\paperlicence}.}}
\vspace*{2mm}

\end{titlepage}


\newpage
\setcounter{page}{2}
\mbox{~}
%
%
%
%

%% file: body.tex
\section{Introduction}
\label{sec:introduction}

Multi-body hadronic decays of charm particles offer an interesting environment for addressing a variety of phenomena related to the interplay of weak and strong interactions. Typically, the formation of three- and four-body final states proceeds through  resonances as intermediate states, with rich interference patterns which allow the study of light meson spectroscopy, among other effects. This is particularly relevant for the controversial scalar sector, which poses a long-standing puzzle. With more states appearing below 2\gev~\cite{PDG2020} than a naïve quark-antiquark nonet can accommodate,\footnote{Natural units with $\hbar = c = 1$ are used.} there is an ongoing debate on their spectra and nature 
\cite{Achasov_4quarknature_2020,Klempt_2007,pelaez2016controversy, Achasov_review2017,Ochs_2013}.

In three-body $D$ decays, the $K\Kb$, $K\pi$ or $\pi\pi$  scattering amplitudes in the  S-wave can be studied in the range from the corresponding threshold up to $\sim$1.5--1.8\gev, thus providing information on meson-meson interactions that are complementary to those from scattering experiments. For the latter, the primary source of information on $L=0$, $I=0$ $\pi\pi$ amplitudes, in particular, comes from the reactions $\pi N\to \pi\pi N$, for $\pi\pi$ invariant mass above 0.6\gev~\cite{Hyams1973,Hyams1975,Protopopescu_pipiscatt,Grayer_pipiscatt,CERN-Cracow-Munich_pipiscatt}, and \mbox{$\Kp \to \pim\pip\ep\nu_e$} decays~\cite{Rosselet_Ke4,BNL-E865_Ke4}, below 0.4\gev. During the last two decades, the increasingly large \D-decay data sets available from charm- and \B-factories have provided new inputs to this field. Currently, the LHCb experiment has the largest samples of charm decays. 

Decays of \D  mesons proceed 
through the weak decay of the charm quark together with the hadronisation forming the final-state mesons, including rescattering processes, before reaching the detector. Scattering amplitudes are, therefore, embedded in the total $D$-decay amplitude, and cannot be directly accessed. However, given that in general most of the decay rates of $D$ mesons are accounted for by resonances, it is often assumed that the dynamics of the final states can be well represented as a quasi-two-body process (``2+1'' approximation \cite{bediagagobel}) and are driven by meson-meson interactions. Nevertheless, effects of coupled channels, three-body interactions and the isospin degree of freedom may also play a role.
At present, there are no tools for a complete description of such decay amplitudes from first principles. 
In this context, the study of the  S-wave in different decay modes -- many with broad and overlapping states for which the ``2+1'' approach is limited -- may provide valuable inputs for phenomenological analyses.

\begin{figure}[h] 
 \centering
\includegraphics[width= 7 cm]{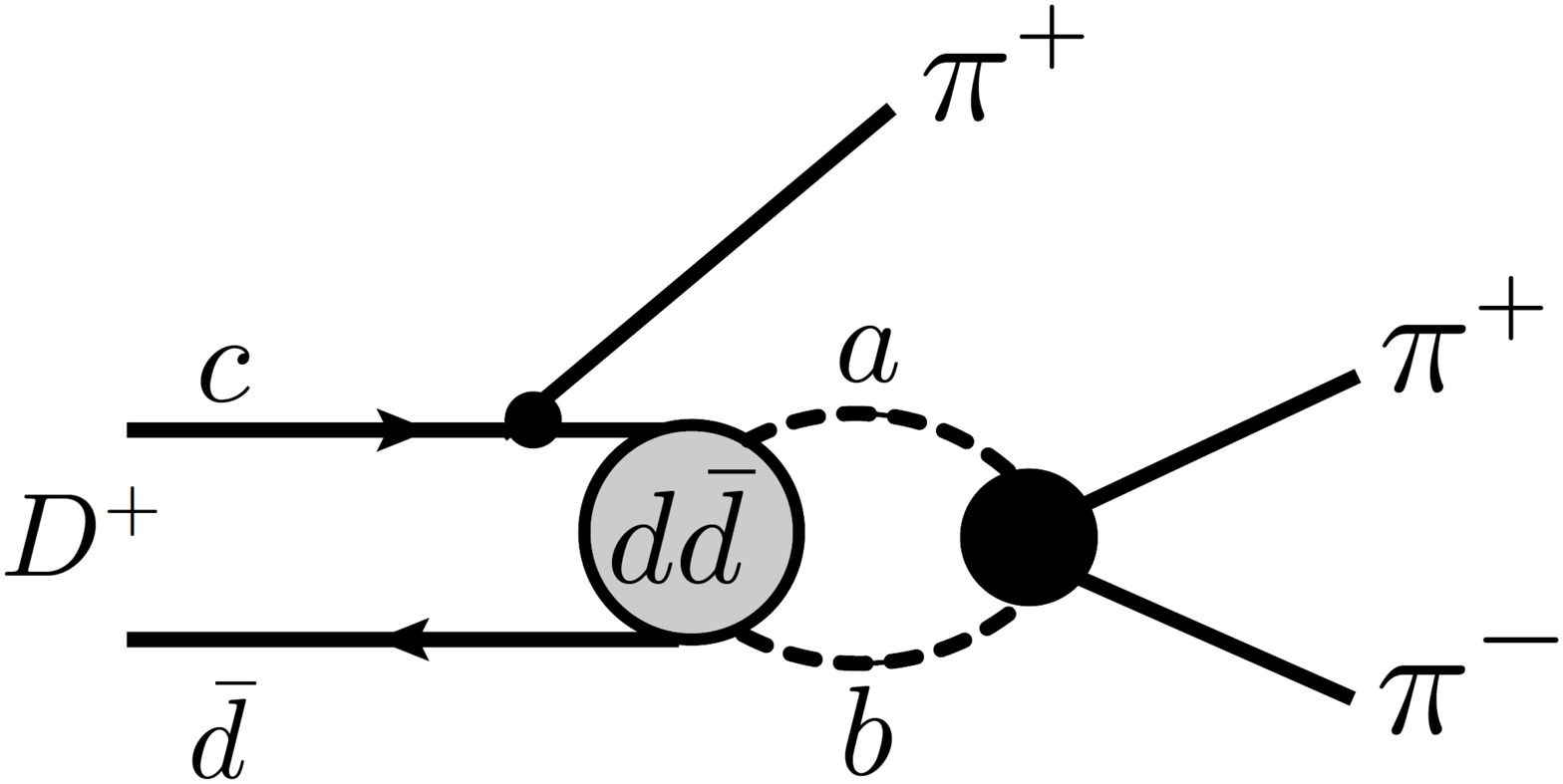}
\caption{Leading diagram for the \Dppp decay. Possible rescattering effects are represented by the $a, b$ meson pair.}
\label{fig:diags0}
\end{figure}

The \Dppp decay\,\footnote{Charge conjugation is implicit throughout this paper.} is a valuable channel for studies of \pim\pip interactions. Previous analyses of this decay~\cite{e791Dp3pi,focus3pi,CleoD3pi} have shown a dominance of the  S-wave component, with a $\sim 50\%$ contribution from the $f_0(500)$ meson. The dominant decay mechanism is expected to be the tree-level external \W-radiation amplitude, illustrated in Fig.~\ref{fig:diags0}, thus the \pim\pip amplitudes, including the  S-wave one, are produced primarily from a $d \bar d$ source. 

As a comparison, in  the \Dsppp decay the  S-wave contribution (produced via  $s \bar s$)  is also found to be dominant \cite{BaBar_Ds3pi,BESIII_Ds3pi,LHCb-PAPER-2022-030}, but its main component is the $f_0(980)$ state, with no evidence for  $f_0(500)$ production.

In this paper, the Dalitz plot of the \Dppp decay is analysed, based on 1.5\invfb of $pp$ collision data at 8\tev centre-of-mass energy, collected by the LHCb experiment in 2012. The main purpose of this work is to determine the resonant structure of the decay and to study  the $\pim\pip$   S-wave  using a {\it quasi-model-independent} partial wave analysis (QMIPWA)  \cite{E791-PWA} for the first time in the $D^+ \to \pi^-\pi^+\pi^+$ final state: no model is assumed for the  S-wave amplitude, which is parameterised as a generic complex function to be determined by a fit to the data, while spin-1 and spin-2 states are included through an isobar model. 

\section{LHCb detector and simulation}
\label{sec:detector}

The LHCb detector \cite{LHCb_detector2008,LHCb-DP-2014-002} is a single-arm forward spectrometer covering the pseudorapidity range $2 < \eta < 5$, designed for the study of particles containing $b$ or $c$ quarks. The detector includes a high-precision tracking system consisting of a silicon-strip vertex (VELO) detector surrounding the $pp$ interaction region, a large-area silicon-strip detector located upstream of a dipole magnet with a bending power of about 4 Tm, and three stations of silicon-strip detectors and straw drift tubes placed downstream of the magnet. The tracking system provides a measurement of the momentum, $p$, of charged particles with a relative uncertainty that varies from 0.5$\%$ at low momentum to 1.0$\%$ at 200\gev. The minimum distance of a track to a primary $pp$ collision vertex (PV), the impact parameter (IP), is measured with a resolution of $(15 + 29/\pt)\mum$, where \pt is the component of the momentum transverse to the beam, in\gev. Different types of charged hadrons are distinguished using information from two ring-imaging Cherenkov detectors. Photons, electrons and hadrons are identified by a calorimeter system consisting of scintillating- pad and preshower detectors, an electromagnetic and a hadronic calorimeter. Muons are identified by a system composed of alternating layers of iron and multiwire proportional chambers.

The online event selection is performed by a trigger system\cite{LHCb-DP-2012-004}, 
which consists of a hardware stage, followed by a software stage, which applies a full event
reconstruction. At the hardware-trigger stage, events are selected based on information from particles that are not related to the signal by requiring a hadron, photon or electron with high transverse energy in the calorimeter, or a muon with high \pt in the muon system.
The software trigger  is divided into two parts. The first part employs a partial reconstruction of the tracks, and a requirement on \pt and IP is applied to, at least, one of the final-state particle forming the \Dp candidate. In the second part  a full event reconstruction is performed and dedicated algorithms are used to select $D^+$ candidates decaying into three charged hadrons.

Simulation is used to model the effects of the detector acceptance and the selection requirements, to validate the fit models and to evaluate efficiencies. In the simulation, $pp$ collisions are generated using \pythia~\cite{Sjostrand:2007gs,Sjostrand:2006za} with a specific \lhcb configuration~\cite{LHCb-PROC-2010-056}. Decays of unstable particles are described by \evtgen~\cite{Lange:2001uf}, in which final-state radiation is generated using \photos~\cite{davidson2015photos}. The interaction of the generated particles with the detector, and its response, are implemented using the \geant toolkit~\cite{Allison:2006ve, Agostinelli:2002hh} as described in Ref.~\cite{LHCb-PROC-2011-006}. 
 For part of the total simulation sample, the underlying $pp$ interaction is reused multiple times, with an independently generated signal decay for each~\cite{LHCb-DP-2018-004}.

\section{\boldmath{\Dppp} selection}
\label{sec:selection}

In the offline selection, combinations of three charged particles with net charge $\pm1$ must form a good-quality vertex well detached from any PV. The PV associated to the \Dp candidate is chosen as that with the smallest value of \chisqip, where \chisqip\ is defined as the difference in the vertex-fit \chisq of the PV reconstructed with and without the particle under consideration, in this case the \Dp candidate. Requirements are placed on the following variables: the significance of the distance between the PV and the \Dp decay vertex; the angle between the reconstructed \Dp momentum vector and the vector connecting the PV to the decay vertex; the $\chi^2$ of the \Dp decay vertex fit;  the distance of closest approach between any two final-state tracks;  the momentum, the transverse momentum and the \chisqip of the \Dp candidate and of its decay products. The invariant mass of the \Dp  candidates,  calculated using the pion mass hypothesis for the three tracks, is required to be within the interval [1800,1940]\mev.

Particle identification (PID) is also used to separate pions from kaons and  muons. The requirements, placed on all tracks, reduce to the percent level the cross-feed from decays such as $D^0\to K^-\pi^+$ plus an unrelated track, $D^+\to K^-\pi^+\pi^+$ and $D^+\to \mu^-\pi^+\pi^+\overline{\nu}_{\mu}$. The $D^+\to \KS\pi^+$ decay is removed by vetoing candidates with $\pi^+\pi^-$ invariant mass in the interval [0.485,0.500]\gev. 

A multivariate analysis (MVA) \cite{TMVA4} is used to further reduce the remaining backgrounds, mostly from random association of three charged tracks, but also from contamination of $D^+_{(s)}\to (\eta^{(')}\to \pi^+\pi^-\gamma)\pi^+$ decays, where the photon is not detected. The MVA uses a Gradient Boosted Decision Tree (BDTG)~\cite{breiman1984classification} as a classifier and is based on the  quantities (or combinations of them)  described above, except PID and  those that can potentially cause large efficiency variation across the Dalitz plot, such as the \chisqip and \pt of the decay products. For training, simulated \Dppp decays are used to represent the signal, whereas data from the invariant-mass sidebands ([1810,1830] and [1910,1930]\mev) of the signal peak are used for the background. Before performing the MVA, a weighting procedure~\cite{rogozhnikov2016reweighting} is used to account for small differences between the simulation and data, including differences in the momentum and transverse momentum distributions of the decay products. The data are divided into two independent sets in a pseudorandom manner.
The $\pim\pip\pip$ mass spectrum of one of the datasets is fitted using the sPlot technique~\cite{Pivk:2004ty} to obtain signal and background weights needed to determine the distributions of the quantities to be used in the MVA, as well as the kinematic distributions, needed for a weighting procedure explained above. The second dataset, corresponding to an integrated luminosity of about 0.75\invfb,  is used to select the final candidates for the Dalitz plot analysis.

The requirement on the BDTG output is chosen to yield a sample of \Dppp decays with 95\% purity in order to reduce the impact of systematic effects related to the background modelling in the Dalitz plot fit. The efficiency drops very rapidly for more stringent requirements, with only a modest gain in purity.

\begin{figure}[h]
\centering
\includegraphics[width=0.6 \textwidth]{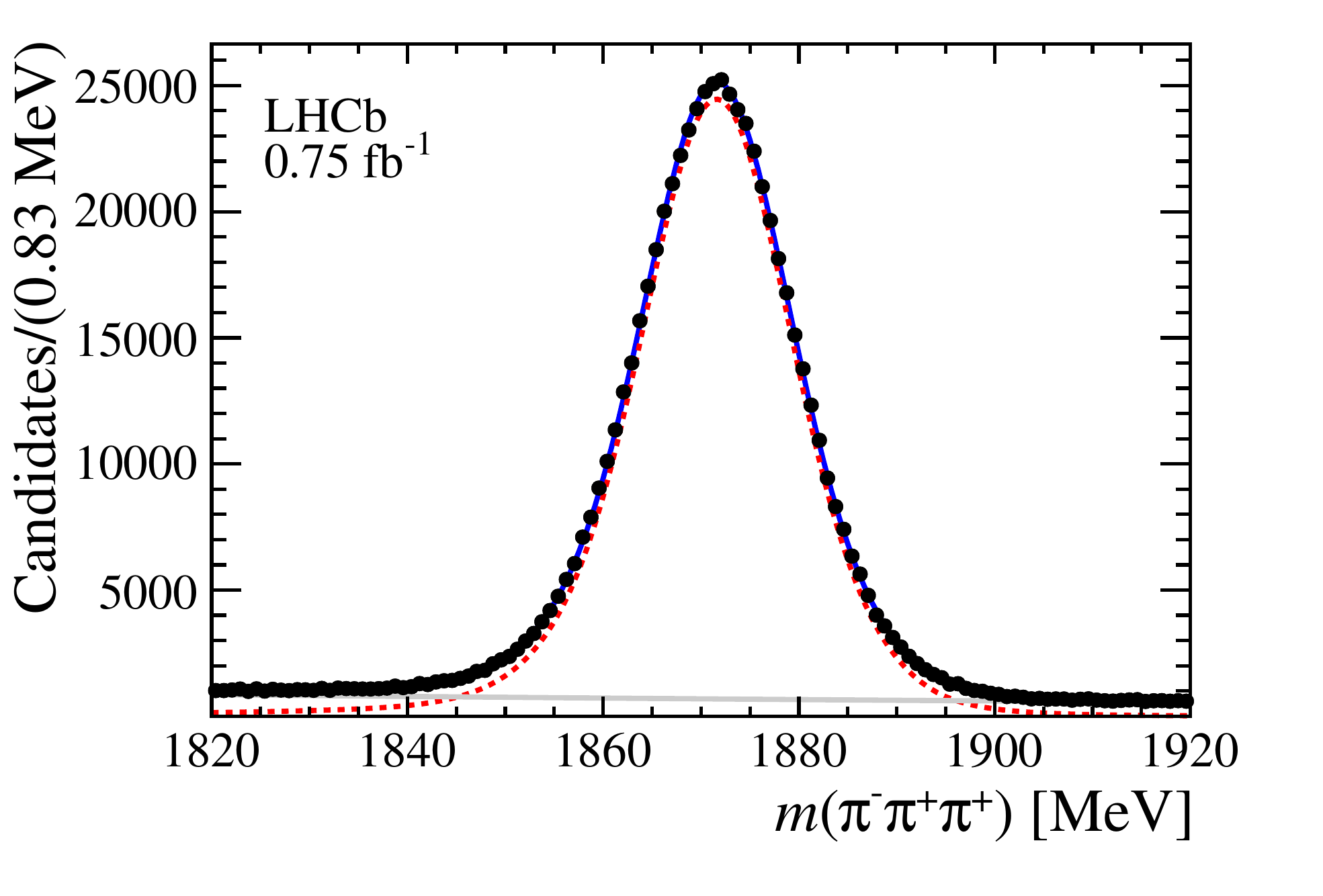}
\caption{Invariant-mass distribution of  \Dppp candidates after final selection, with the fit result superimposed (blue solid line). The dashed red line and the solid gray line correspond to the signal and background components of the fit, respectively.
}
\label{fig:final_3pimass}
\end{figure}

The invariant mass of  \Dppp candidates after all requirements is shown in Fig.~\ref{fig:final_3pimass}. An extended binned maximum-likelihood fit to this distribution is performed.
The probability distribution function (PDF) of the signal  
is represented by a sum of a Gaussian function and two Crystal Ball (CB)~\cite{Skwarnicki:1986xj} functions, while the background is modelled by
an exponential function. The signal PDF is
\begin{multline}
P_{\rm sig}[m(\pim\pip\pip)] = f_{\rm G} \times {\rm G}(\mu, \sigma_{\rm G}) + (1-f_{\rm G})\times
\left[ f_{\rm CB} \times {\rm CB}_1(\mu, R_{1}\sigma_{\rm G},\alpha_1,N_1) \, + \right. \\
\left.  (1-  f_{\rm CB})\times {\rm CB}_2(\mu, R_{2}\sigma_{\rm G},\alpha_2,N_2)\right], 
\label{spdf}
\end{multline}
\noindent where $\mu$ and $\sigma_{\rm G}$ are the mean value and the width of the Gaussian function G. The two Crystal Ball functions, CB$_1$
and CB$_2$, have widths  $R_{1}\sigma_{\rm G}$ and  $R_{2}\sigma_{\rm G}$, and tail parameters $\alpha_1$, $N_1$ and $\alpha_2$,  $N_2$. A common 
parameter, $\mu$, describes the most probable mass value of the two Crystal Balls and the mean of the Gaussian function. The fractions of each PDF component are $f_{\rm G}$ for the Gaussian function, $(1-f_{\rm G})\times f_{\rm CB}$ for CB$_1$ and $(1-f_{\rm G})\times(1-f_{\rm CB})$ for CB$_2$.
The parameters $\alpha_i$, $N_i$, $R_{i}$, $f_{\rm CB}$ and $f_{\rm G}$ defining the signal PDF are fixed to the values obtained from a fit to the simulated sample, while the parameters $\mu$ and $\sigma_G$ are allowed to float freely. 

For the Dalitz Plot analysis, candidates are selected within a $2\sigma_{\rm eff}$ mass window around the mean $\mu$, where the effective mass resolution, 
$\sigma_{\rm eff}$, is defined by 
\mbox{$\sigma_{\rm eff}=\sqrt{\left[ f_{\rm G} +(1-f_{\rm G})f_{\rm CB} R_1^2 + (1-f_{\rm G})(1-f_{\rm CB})R_2^2 \right] \sigma_{\rm G}^{2} }=8.7\mev$}.
The final sample has 601\,171 candidates with a signal purity of $(95.2 \pm 0.1)\%$ in the corresponding mass interval of [1854.1,1889.0]\mev. 
Multiple candidates in the same event correspond to only 0.15\% of the final sample and are retained.

\begin{figure}[h]
\centering
\includegraphics[width=0.6\textwidth]{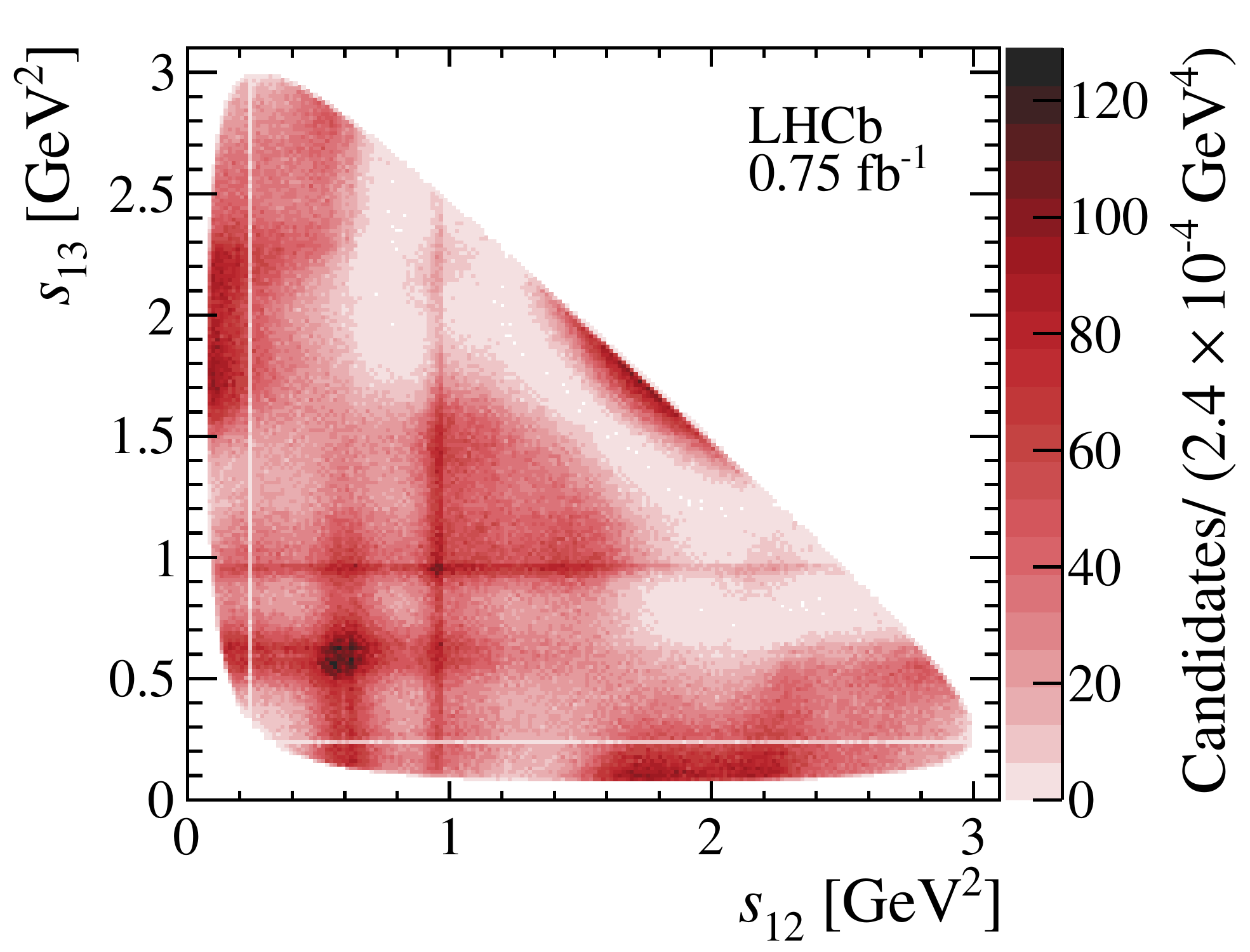}
\caption{Dalitz plot distribution of the final \Dppp sample. The lines in the interval $[0.235,\!0.250]\gev^2$ correspond to the veto applied to remove $\Dp \to \KS \pip$ decays.}
\label{fig:final_Dalitzplot}
\end{figure}

The Dalitz plot distribution of the selected candidates is shown in Fig.~\ref{fig:final_Dalitzplot}. The axes are the Dalitz variables $s_{12}\equiv (p_1+p_2)^2$ and 
$s_{13}\equiv (p_1+p_3)^2$, where $p_i$ ($i=1,2,3$) are the 4-momenta of the three pions in the final state and the particle ordering is such that the pion with charge  opposite to that of  the $D^+$ meson
is always particle 1, 
and the same-sign pions are randomly assigned particles 2 and 3. These Lorentz-invariant quantities are 
computed after refitting tracks with the constraint on the invariant mass of the candidate to the known \Dp mass~\cite{PDG2020}.

\section{Background and efficiency models}
\label{backeff}
\begin{figure}[h]
    \centering
    \includegraphics[width = 0.495\textwidth]{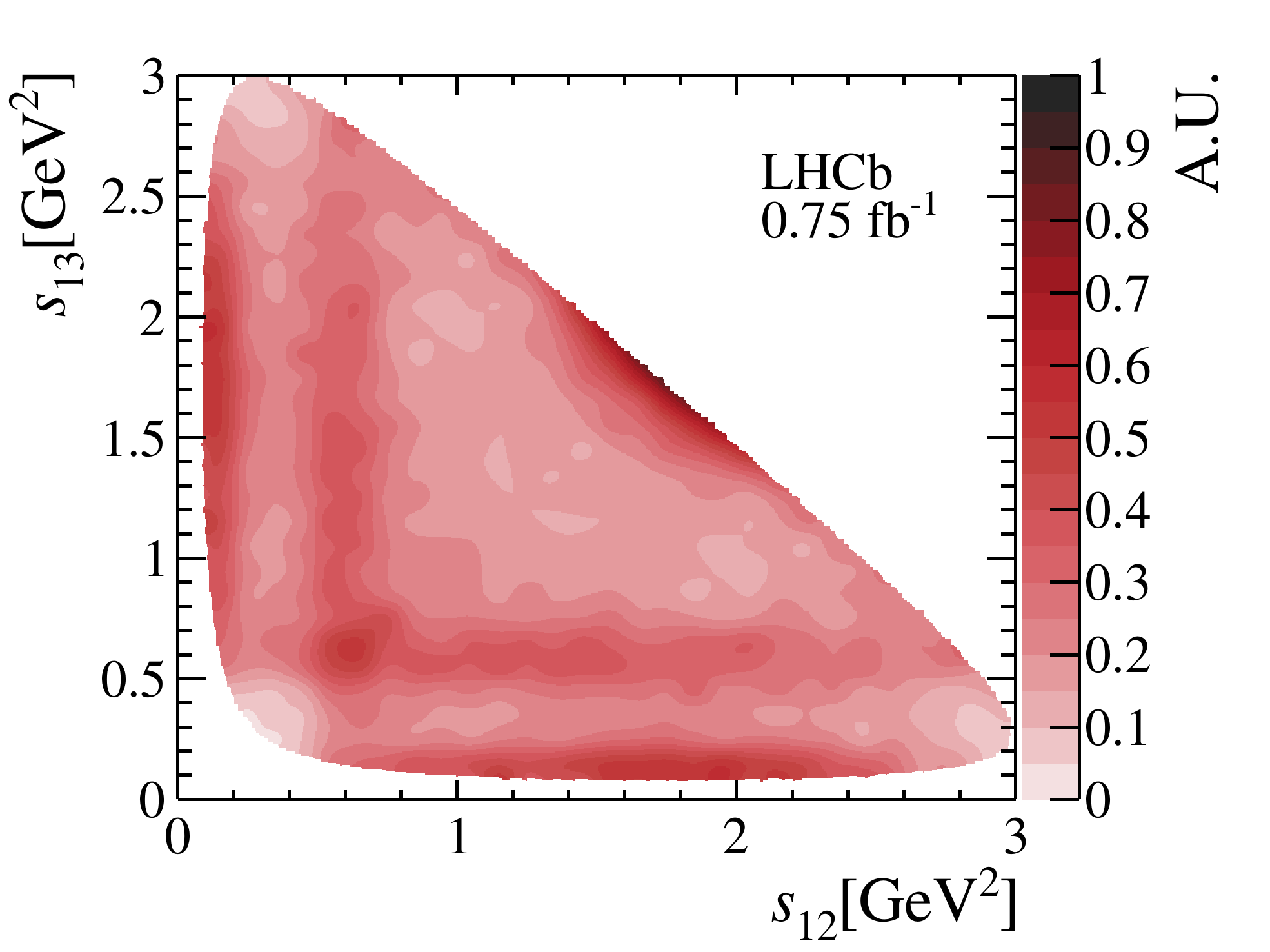}
    \includegraphics[width = 0.495\textwidth]{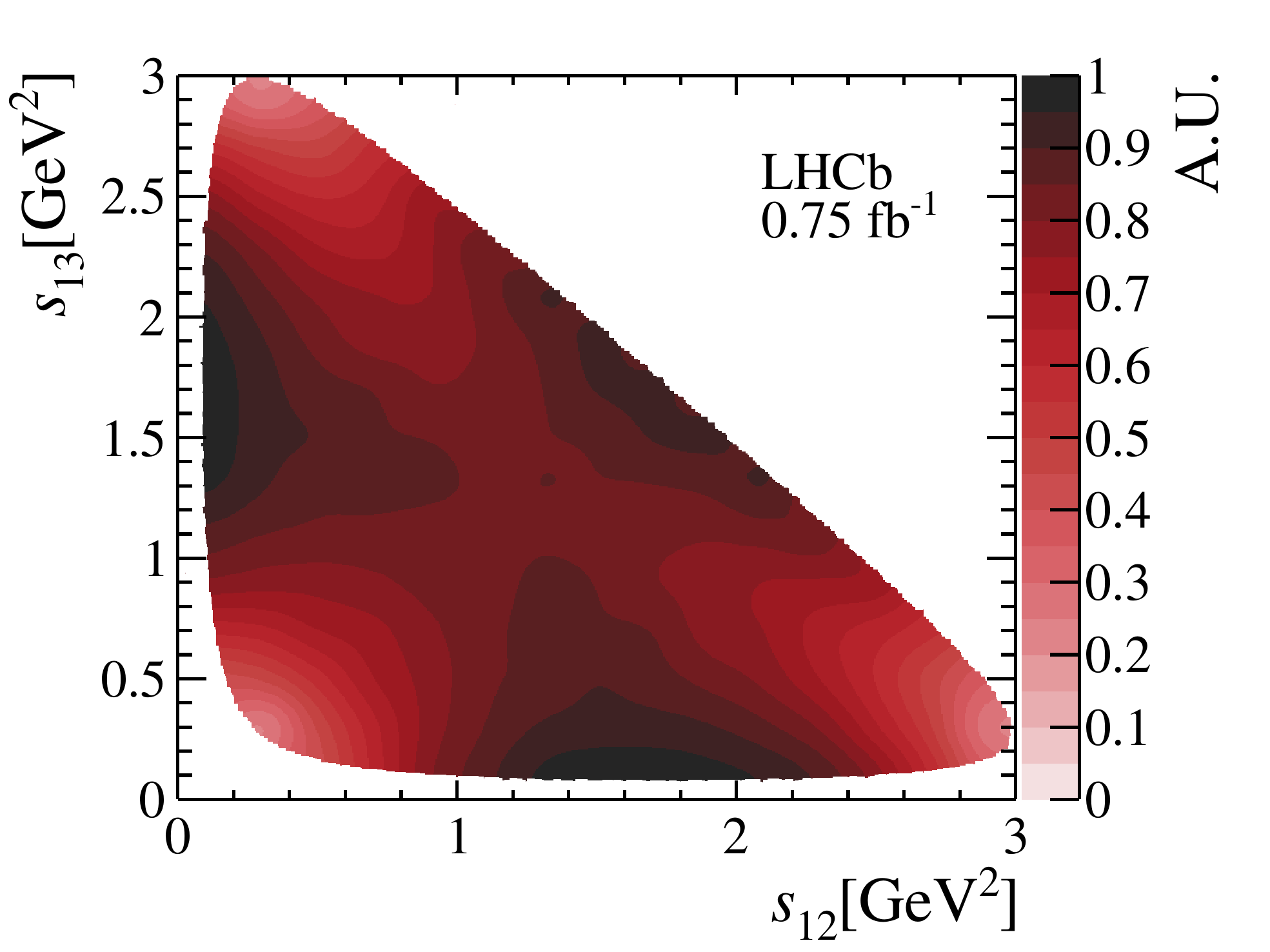}
    \caption{Models for (left) background distribution and (right) signal efficiency across the Dalitz plot, where the $z$-axis scale is arbitrary.}
    \label{fig:back_eff}
\end{figure}

The remaining background in the \Dppp sample, which amounts to 4.8\%, needs to be parameterised to be included in the Dalitz plot fit. The background distribution is inferred from the sidebands of the $D^+ \to \pi^- \pi^+ \pi^+$ invariant-mass signal region, specifically, the intervals [1810,1830]\mev and [1910,1930]\mev. The background within the signal region is assumed to be the average composition of both sidebands. 
The candidates in these regions are projected to the Dalitz plot and a two-dimensional (2D) cubic spline procedure~\cite{laura++} is used to smooth the distribution in order to avoid binning discontinuities. The resulting model, shown in Fig.~\ref{fig:back_eff}~(left),  is used in the Dalitz plot fit.  

The signal distribution in the Dalitz plot includes efficiency effects, which need to be corrected for. The efficiency model across the Dalitz plot includes the effects of the geometrical acceptance of the detector, as well as reconstruction, trigger, selection, and PID requirements. All these effects apart from those associated with PID are quantified using simulation. The PID efficiencies for the pions are evaluated from calibration samples of $D^{*+} \to D^0(\to K^-\pi^+)\pi^+$ 
decays~\cite{LHCb-PUB-2016-021} and depend on the particle momentum, pseudorapidity and 
charged-particle multiplicity.  
The final efficiency model is constructed from a two-dimensional histogram with $15\times 15$ uniform bins which is then smoothed by a 2D cubic spline, as shown in Fig.~\ref{fig:back_eff}~(right).

\section{The QMIPWA  formalism}
\label{sec:formalism}

The Dalitz plot of \Dppp candidates shown in Fig.~\ref{fig:final_Dalitzplot} shows a rich resonant structure: contributions from the subchannels $\rho(770)^0\pip$ (with an evident interference with $\omega(782)\pip$),  $f_0(500)\pip$, $f_0(980)\pip$ and  possible high-mass vector and tensor states are seen. 
To disentangle these contributions, a full amplitude analysis is needed. The approach of this work is to describe the total \Dppp amplitude as a coherent sum of an  S-wave contribution and higher-spin waves, 
 \begin{equation}
     \mathcal{A}\left(s_{12}, s_{13}\right)= \left[ \mathcal{A}_{\rm S\mbox{-}wave}(s_{12})+\sum_{i} a_{i} e^{i \delta_{i}}\mathcal{A}_i\left(s_{12}, s_{13}\right) \right]+\left(s_{12} \leftrightarrow s_{13}\right),
     \label{eq:fit}
 \end{equation}
\noindent where the total amplitude is Bose-symmetrised with respect to $s_{12}$ and $s_{13}$ due to the two identical pions. 
The first term is the  S-wave amplitude, 
\begin{equation}
    \mathcal{A}_{\rm  S\mbox{-}wave}(s_{12}) = a_0(s_{12}) e^{i\delta_0(s_{12})},
\end{equation}
where the real functions $a_0(s_{12})$ and $\delta_0(s_{12})$ are to be determined by the Dalitz plot fit: the $\pim\pip$ invariant-mass spectrum  is divided in 50 intervals (knots) where interpolation to obtain a continuous  S-wave complex function is attained through a linear spline. 
The intervals are not uniformly distributed to allow better determination of the magnitude and phase variations in regions where they are expected to vary most -- such as near the $f_0(980)$ resonance. Variations in the knot spacing and number of knots are addressed for systematic uncertainties. 
The P- and D-wave components are included through an isobar model, represented by the  terms in the sum in Eq.~\ref{eq:fit}, where $\mathcal{A}_i\left(s_{12}, s_{13}\right)$ is the complex amplitude of resonance $R_i$, with magnitude $a_i$ and phase $\delta_i$ as free parameters. This approach to the total amplitude is referred to as {\it quasi-model-independent} since any limitation of the isobar model to describe the higher-spin components may reflect in the description of the  S-wave amplitude which has 100 free parameters.

Within the isobar model, the individual resonant amplitude  for a process of type $D \rightarrow R_i \,c$;$R_i \rightarrow a \,b$, where $R_i$ is an intermediate resonant state, is written as
\begin{eqnarray}
\mathcal{A}_{i}\left(s_{ab}, s_{ac}\right)=F_D^J \, F_R^J \,  \mathcal{M}_J \,T_{R_i}(s_{ab}).
\label{eq:ai_amp}
\end{eqnarray}
Since the $D^+$ meson and the final-state pions are spinless particles, the spin of the resonance, $J$, is equal to the orbital angular momentum in both the decays $D \rightarrow R_i \, c$ and $R_i \rightarrow a\, b$.

The Blatt-Weisskopf barrier factors \cite{book:BlattWeisskopf}, $F_D^J$ and $F_R^J$, take into account the finite size of the $D$ meson and the $R_i$ resonance in the decay processes $D\to R_i \, c$ and $R_i\to a\, b$, respectively. They are functions of $z=rp^*$, as shown in Table~\ref{table:blatt}, where $r$ is the effective radius of the decaying particle, $p^*$ is the modulus of the momentum of the decay products measured in the decaying particle rest frame, and $z_0$ is the value of $z$ calculated at the known mass of the decaying particle. In this analysis, the effective radii of the $D^+$ meson and the intermediate resonances are set to $5$ GeV$^{-1}$ and $1.5$ GeV$^{-1}$, respectively \cite{e791Dp3pi,Cleo-Dalitz-Kpipi,LHCb-PAPER-2018-039,LHCb-PAPER-2021-045}.

The function $\mathcal{M}_J$ describes the angular distribution of the decay particles in the Zemach formalism~\cite{paper:Zemach} with explicit forms for $J=1$ and $J=2$ given by
\begin{equation}
\mathcal{M}_{1} = s_{bc} - s_{ab} + \left(\frac{1}{s_{ab}} (m_D^2 - m_c^2) (m_a^2 - m_b^2)\right),
\label{spin1}
\end{equation}
and
\begin{multline}
\mathcal{M}_{2}= \mathcal{M}_{1}^2 - \frac{1}{3}\left(s_{ab} - 2m_D^2 -m_c^2 + \frac{1}{s_{ab}}(m_D^2-m_b^2)^2\right)\times \\ \left(s_{ab} - 2m_a^2 - 2m_c^2 + \frac{1}{s_{ab}}(m_a^2-m_b^2)^2 \right),
\label{spin2}
\end{multline}
respectively, where $m_D$ is the known mass of the $D$ meson and $m_a$, $m_b$ and $m_c$ those of the  decay products. It is clear that these expressions take a simpler form for the particular case when there are three charged pions in the final state ($m_a=m_b=m_c=m_\pi$).

\begin{table}[t] 
    \caption{Spin-dependent Blatt-Weisskopf barrier factors. }
	\begin{center}
	\begingroup
		\begin{tabular}{ l  c }
Spin		 & Blatt-Weisskopf factor  \\
		\hline \\[-0.2cm]
			$J=1$ & $\sqrt{\dfrac{1 + z_0^2}{1+z^2}}$  \\[0.5cm]
			$J=2$ & $\sqrt{ \dfrac{z_0^4 + 3z_0^2 + 9}{z^4 + 3z^2 + 9}}$   \\[0.5cm]
			\hline  
		\end{tabular}
		\endgroup
	\end{center}
	\label{table:blatt}
\end{table} 
The dynamical function $T_{R_i}(s_{ab})$ in Eq.~\ref{eq:ai_amp} represents the resonance lineshape, usually a relativistic Breit-Wigner (RBW),
\begin{eqnarray}
T_{\rm RBW}(s_{ab})=\frac{1}{m_{0}^{2}-s_{ab}-i m_{0} \Gamma(s_{ab})},
\end{eqnarray}
where  $m_0$ is the nominal mass of the resonance and  $\Gamma(s_{ab})$ is the mass-dependent width which is given by 
\begin{eqnarray}
\Gamma\left(s_{ab}\right)=\Gamma_{0}\frac{m_{0}}{\sqrt{s_{ab}}}\left(\frac{p^*}{p^*_{0}}\right)^{2 J+1}  (F_{R}^{J})^{2}\, ,
\end{eqnarray}
with $\Gamma_0$ being the nominal width of the resonance. 

A Gounaris-Sakurai (GS) function~\cite{gounaris1968finite} is a modification of the RBW lineshape,  commonly used to describe the pion electromagnetic form factor in the parameterisation of spin-1 $\rho$-type resonances, 
\begin{equation}
T_{\rm GS}(s_{ab}) =  \frac{1+\Gamma_0d/m_0}{(m_0^2 - s_{ab}) + f(s_{ab}) - i m_0 \Gamma(s_{ab})}, 
\end{equation}
where
 \begin{equation}
\begin{split}
f(s_{ab}) &= \Gamma_0 \,\frac{m_0^2}{p_0^{*3}}\,\left\{\; p^{*2} \left[ h(s_{ab})-h(m_0^2)\right] +
       \left(\,m_0^2-s_{ab}\,\right)\,p^{*2}_0\,
       \frac{\rm{d}h}{\;~ {\rm d}s_{ab}}\bigg|_{m_0^2}
       \;\right\}. \\
    \end{split}
\end{equation}
\noindent The function $h(s_{ab})$ is given by
\begin{equation}
    h(s_{ab}) = \frac{2}{\pi}\,\frac{p^*}{\sqrt{s_{ab}}}\,
       \ln\left(\frac{\sqrt{s_{ab}}+2p^*}{2m_\pi}\right)~, \\
\end{equation}
where $m_{\pi}$ is the pion mass and the derivative is given by
\begin{equation}
\frac{{\rm d}h}{\;~ {\rm d}s_{ab}}\bigg|_{m_0^2} =
h(m_0^2)\left[(8p_0^{*2})^{-1}-(2m_0^2)^{-1}\right] \,+\, (2\pi m_0^2)^{-1}~. \\
\end{equation}
\noindent The parameter $d=f(0)/(\Gamma_0 m_0)$ is given by
\begin{equation}
  d = \frac{3}{\pi}\frac{m_\pi^2}{p_0^{*2}}
  \ln\left(\frac{m_0+2p^*_0}{2m_\pi}\right)
  + \frac{m_0}{2\pi\,p^*_0}
  - \frac{m_\pi^2 m_0}{\pi\,p_0^{*3}}~.
\end{equation}

In this analysis, the GS lineshape is used for the spin-1 $\rho$-type resonances, while  the RBW lineshape is used for other resonances such as the $\omega(782)$ and  $f_2(1270)$ states. The values $m_0$ and $\Gamma_0$ for all resonances are fixed in the fit to their known values \cite{PDG2020}. 

The $\omega(782)\to \pim\pip$ decay violates isospin, and it is not clear whether this process occurs through direct decay or  through mixing with the $\rho(770)^0$ state (or both). 
As an alternative to representing the $\rho(770)\pip$ and $\omega(782)\pip$ amplitudes as a sum of isobars, their combined contribution is parametrised through a $\rho-\omega$ mixing lineshape given by \cite{Rensing:1993hf}
\begin{equation}
    T_{\rho-\omega}=T_{\rho}\left[\frac{1+ \Delta |B|e^{i\phi_B} T_{\omega}}{1-\Delta^2T_{\rho}T_{\omega}}\right],
    \label{rhomix}
\end{equation}
\noindent where $T_{\rho}$ and $T_{\omega}$ are  the GS and RBW lineshapes for the $\rho(770)^0$ and  $\omega(782)$ resonances, respectively. The magnitude $|B|$ and the phase $\phi_B$ quantify the relative contribution  of the $\omega(782)$ and the $\rho(770)$ resonances, and are free parameters in the fit. The factor $\Delta = \delta (m_0^{\rho}+ m_0^{\omega})$ governs the electromagnetic mixing of these states, where the value of $\delta$ is fixed to $2.15 \mev$ \cite{Rensing:1993hf} and $m_0^{\rho}$ and $ m_0^{\omega}$ are the known masses\cite{PDG2020}. This parameterisation is equivalent to that used in Ref.~\cite{CMD2-2002} given that $\Delta^2$ is small and therefore the term where it appears in the denominator can be neglected.

\section{The Dalitz plot fit methodology}

Given the large data sample, the large number of parameters used in the  decay amplitude, and the need to normalise the total PDF at each iteration of the minimisation process, the GooFit framework \cite{Goofit} for maximum-likelihood fits is used. GooFit is based on GPU acceleration with parallel processing.

An unbinned maximum-likelihood fit to the data distribution in the Dalitz plot is performed. The likelihood function is written as a combination of the signal and background PDFs given by
\begin{eqnarray}
\mathcal{L}=\prod_i \left\{ f_{\rm sig} \times \mathcal{P}^i_{\text {sig}}\left(s_{12}, s_{13}\right)+\left(1-f_{\rm sig}\right) \times \mathcal{P}^i _{\text {bkg}}\left(s_{12}, s_{13}\right) \right\},
\end{eqnarray}
where $f_{\rm sig}$ is the signal fraction and the product runs over the candidates in the final data sample. The background PDF, $\mathcal{P}_{\rm bkg}(s_{12},s_{13})$,  is the normalised background model, and is provided as the high definition histogram shown in Fig.~\ref{fig:back_eff}.
The normalised signal PDF, $\mathcal{P}_{\rm sig}(s_{12},s_{13})$, is given by

\begin{eqnarray}
\mathcal{P}_{\mathrm{sig}}\left(s_{12}, s_{13}\right)=\frac{\epsilon\left(s_{12}, s_{13}\right) \left|\mathcal{A}\left(s_{12}, s_{13}\right)\right|^{2} }
{\iint_{\mathrm{DP}}\epsilon\left(s_{12}, s_{13}\right) \left|\mathcal{A}\left(s_{12}, s_{13}\right)\right|^{2}d s_{12} d s_{13}},
\end{eqnarray}
where  $\epsilon(s_{12}, s_{13})$ is the efficiency model function included as the smoothed histogram shown in Figure~\ref{fig:back_eff}; the denominator is the integral of the numerator over the Dalitz plot (DP) to guarantee that $\mathcal{P}_{\mathrm{sig}}\left(s_{12}, s_{13}\right)$ is normalised at each iteration of the minimisation process.  The fit parameters are the 50 pairs of magnitudes and phases of the  S-wave amplitude, and the magnitudes and phases of the higher-spin components, except the $\rho(770)^0\pip$ channel which is taken as the reference mode, with magnitude fixed to 1 and  phase fixed to zero. The set of optimal parameters is determined by minimising the quantity $-2\log \mathcal{L}$ using the \texttt{MINUIT} \cite{Minuit} package.

The fit fraction for the $i^{\rm th}$ intermediate channel is defined as

\begin{equation}
    {\rm FF}_{i}=\frac{\iint_{\mathrm{DP}}\left|a_{i} e^{i\delta_i} \mathcal{A}_{i}\left(s_{12}, s_{13}\right)\right|^{2} d s_{12} d s_{13}}{\iint_{\mathrm{DP}}\left|\sum_j a_{j} e^{i\delta_j}\mathcal{A}_j\left(s_{12}, s_{13}\right)\right|^{2} d s_{12} d s_{13}}\, .
\end{equation}
Due to interference, the sum of fit fractions can be less than or greater than 100\%. Interference fit fractions can also be defined, quantifying the level  of interference between any pair $i,j ~(i\ne j$) of  amplitude components,
\begin{equation}
    {\rm FF}_{i j}=\frac{\iint_{D P} 2 \operatorname{Re}\left[a_{i}a_{j} e^{i(\delta_i-\delta_j)} \mathcal{A}_{i}\left(s_{12}, s_{13}\right) \mathcal{A}_{j}^{*}\left(s_{12}, s_{13}\right)\right] d s_{12} d s_{13}}{\iint_{\mathrm{DP}}\left|\sum_k a_{k} e^{i\delta_k} \mathcal{A}_k\left(s_{12}, s_{13}\right)\right|^{2} d s_{12} d s_{13}}.
\end{equation}
By construction, the sum of fit fractions and interference terms is 100\%.

The fit quality is measured through the statistical quantity $\chi^2$ defined as
\begin{equation}
    \chi^{2}=\sum_{i=1}^{N_{b}} \chi^{2}_i =\sum_{i=1}^{N_{b}} \frac{(N_i^{\rm obs}-N_i^{\rm est})^2}{\sigma_i^2},
\end{equation}
where the Dalitz plot is divided in $N_b$ bins and, for each bin, the number of observed candidates, $N_i^{\rm obs}$, the number of candidates estimated from the fit model, $N_i^{\rm est}$, and the uncertainty on their difference,  $\sigma_i$, are obtained. 
For unbinned maximum-likelihood fits, the number of degrees of freedom (ndof) range as $[N_b-q-1, N_b-1]$ \cite{MWilliams}, where $q$ is the number of free parameters,  and it is used to calculate the corresponding range of  $\chi^2/\textrm{ndof}$. 
This is done using the {\it folded} Dalitz plot -- due to the symmetry of the \Dppp Dalitz plot with respect to the axis  $s_{13}=s_{12}$, the variables $s_{\rm high}$ and $s_{\rm low}$ are defined, respectively, as the higher and the lower values of each pair $(s_{12},s_{13})$. The folded Dalitz plot is divided in $N_b = 625$ bins using an adaptive binning algorithm, such that all bins have the same population. Besides the $\chi^2/\textrm{ndof}$, the value of $-2\log \mathcal{L}$ is also used to compare models. In addition, the  distribution of residuals $(N_i^{\rm obs}-N_i^{\rm est})/\sigma_i$  across the folded Dalitz plot is used for visual inspection of any local discrepancy between fit model and data, which are also compared through the projections of $s_{\rm high}$, $s_{\rm low}$, the sum of these projections, denoted $s_{\pim\pip}$, and $s_{23} \equiv (p_2+p_3)^2$.

\section{Dalitz plot fit results}
\label{sec:results}

The \Dppp amplitude fit model is constructed with the scalar sector represented through the QMIPWA approach using 50 knots, and starting with the spin-1 and spin-2 states observed from previous analyses of this decay (E791~\cite{e791Dp3pi}, FOCUS~\cite{focus3pi} and CLEO~\cite{CleoD3pi} collaborations, with  much smaller datasets): $\rho(770)^0\pip$,  $\rho(1450)^0\pip$ and $f_2(1270)\pip$, plus the $\omega(782)\pip$ channel, not observed in previous analyses but clearly seen in Fig.~\ref{fig:final_Dalitzplot}. From that,  other possible states, such as $f_2'(1525)$, $\rho(1700)^0$, $\rho_3(1690)$,\footnote{Since the $\rho_3(1690)$ state was not found to be significant, the formalism for spin-3 resonances is not described in  Sec.~\ref{sec:formalism}.} are added one at a time until a good representation of the data is found.

\begin{figure}[h]
\begin{center}
       \includegraphics[width = 0.495\textwidth]{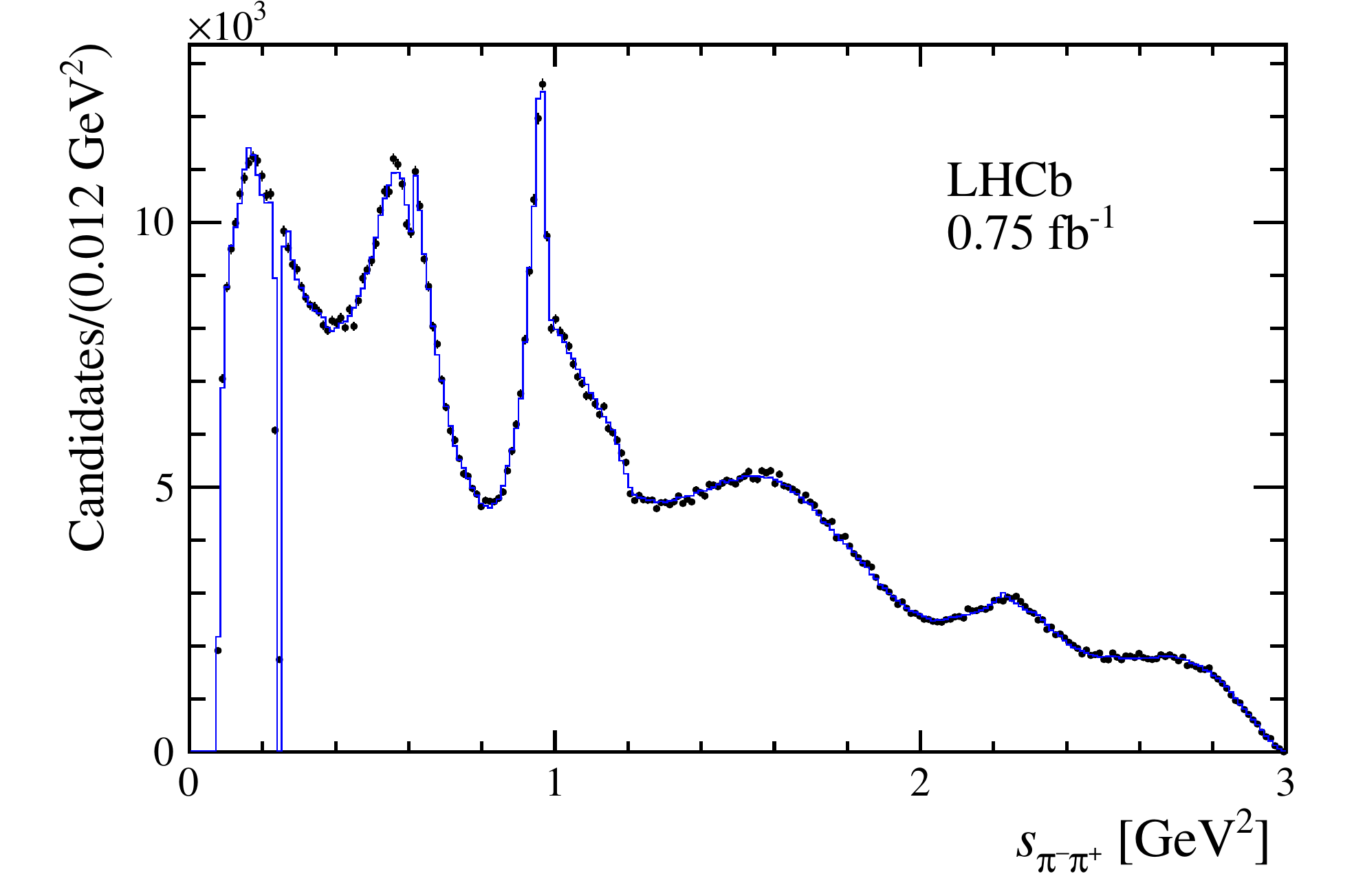}
       \includegraphics[width = 0.495\textwidth]{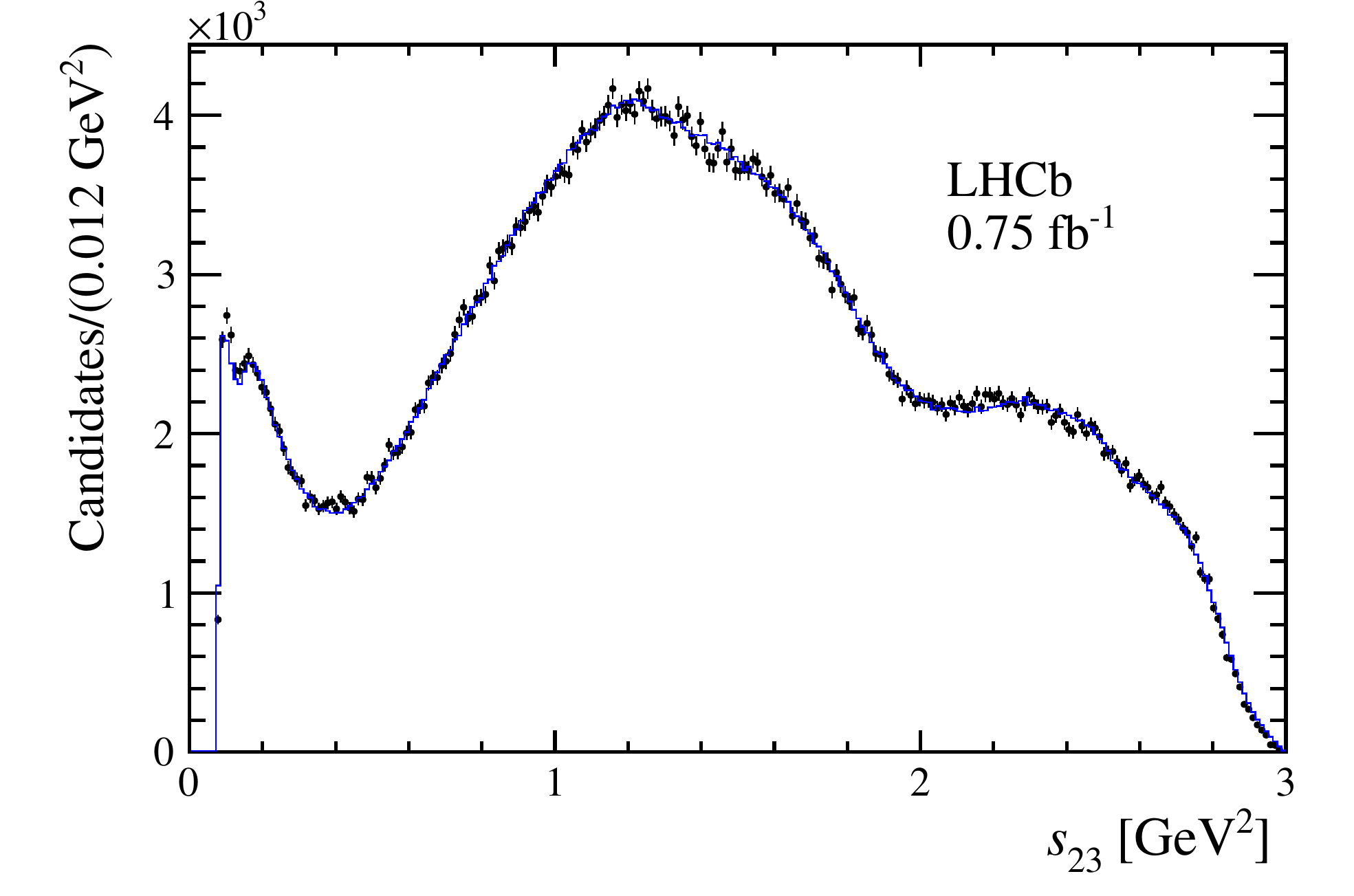}
    
    \includegraphics[width = 0.495\textwidth]{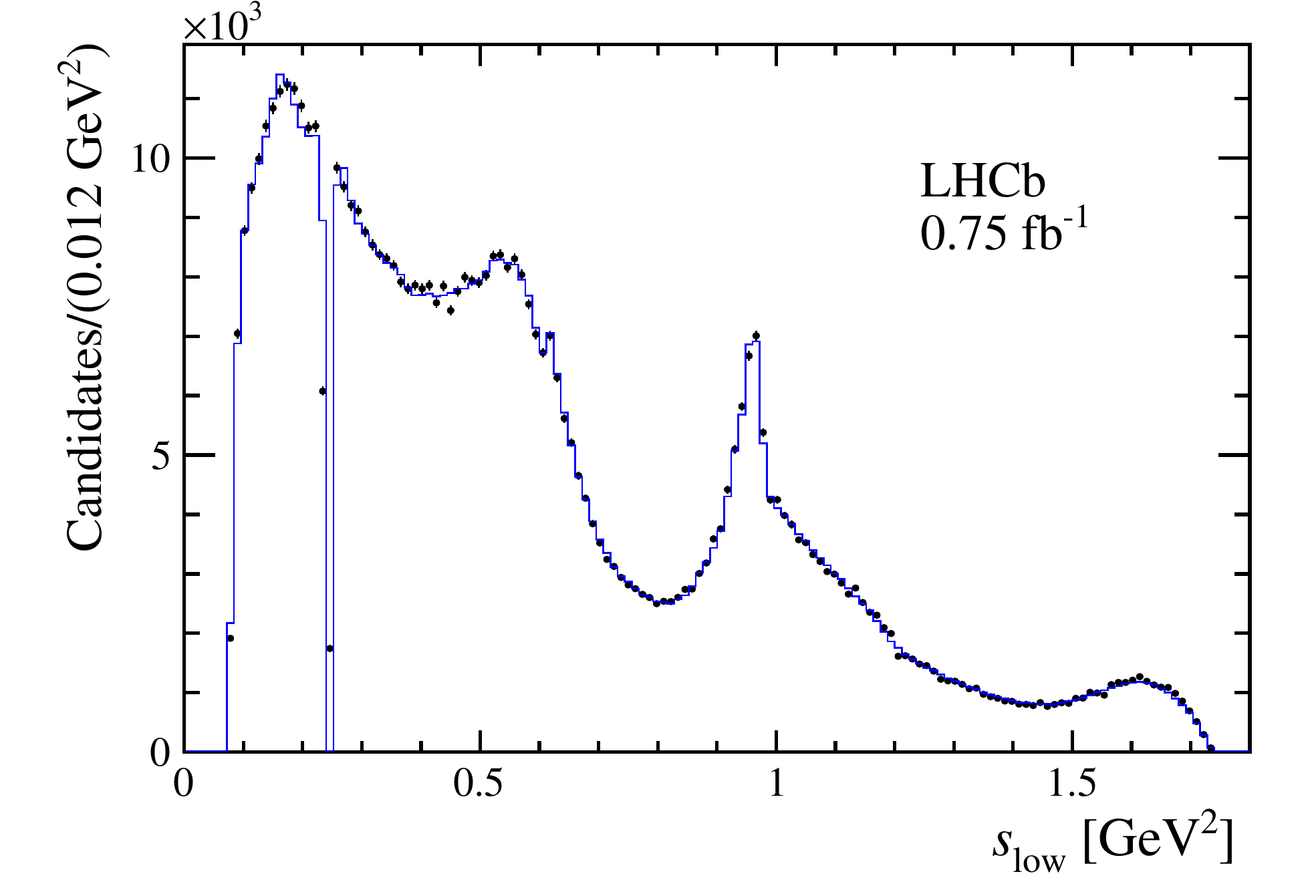}
    \includegraphics[width = 0.495\textwidth]{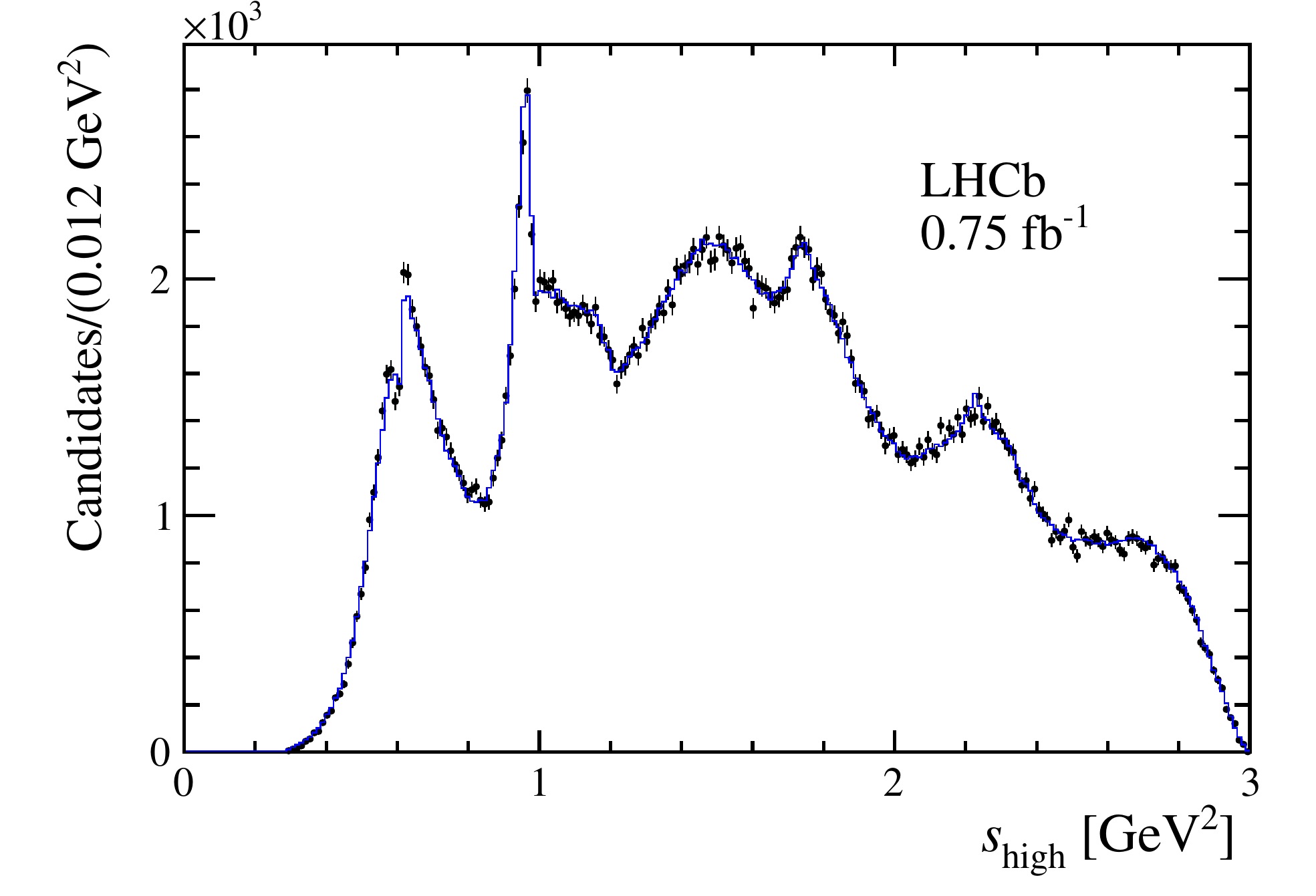}
    
    \includegraphics[width = 0.5\textwidth]{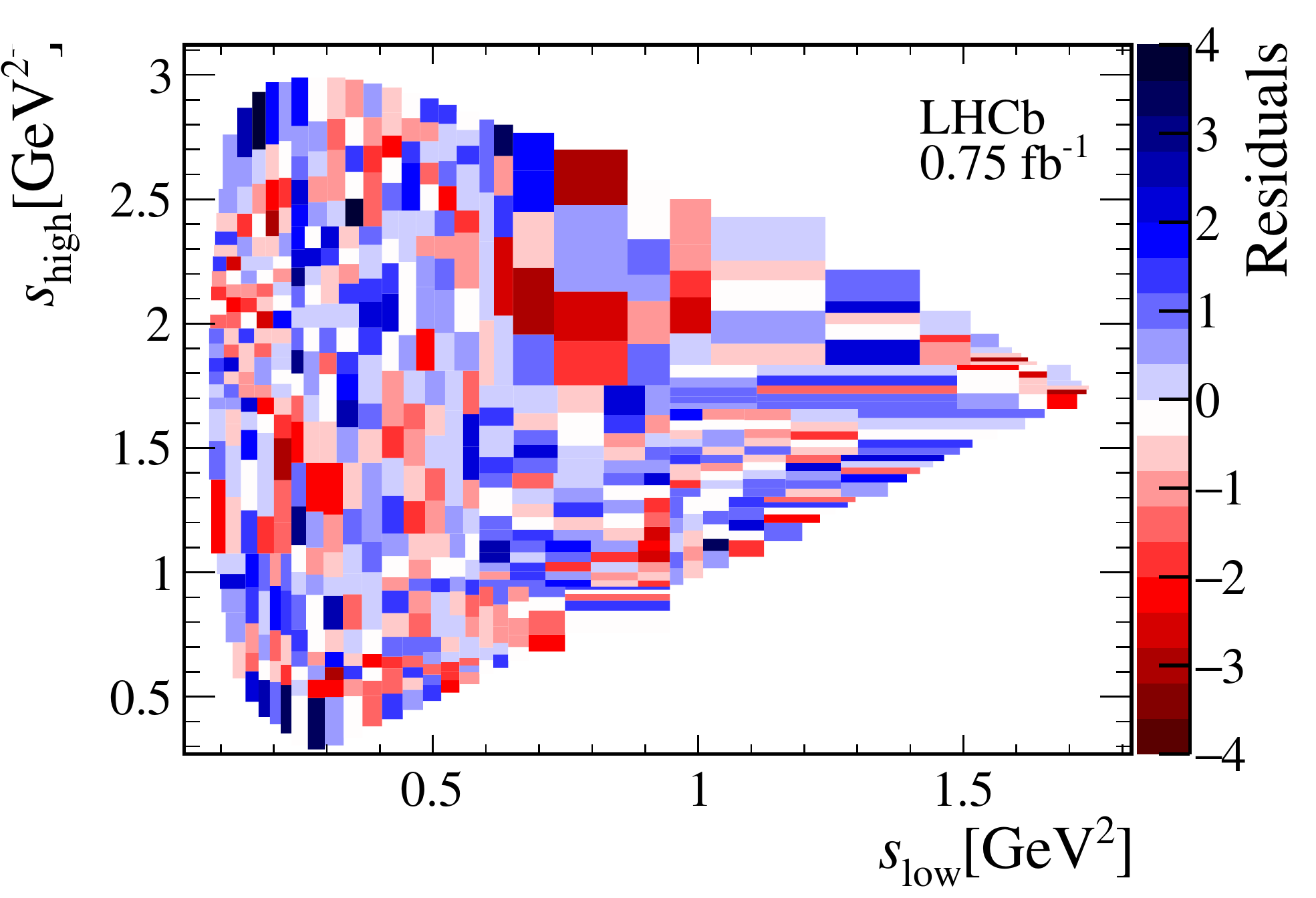}
\end{center}
    \caption{Dalitz plot projections of  (top left) $s_{\pim\pip}$, (top right) $s_{23}$, (middle left) $s_{\rm low}$,  and (middle right) $s_{\rm high}$ projections, where the (red) points are data and the (blue) line is the fit model result, with the fit normalised residuals displayed in the bottom plot. }
    \label{fig:fitPWA3}
\end{figure}

The best model is achieved when the $\rho(1700)^0$ resonance is added;  attempts to include further states do not bring significant improvements. This model has 108 free parameters. 
Table~\ref{tab:finalPWA3} summarises the results from the fit, including systematic uncertainties discussed later in Sec.\ref{sec:systematics}. Interference fit fractions are shown in Table~\ref{tab:intFF}. 
The projections and  the  distribution of residuals are  shown in Fig.~\ref{fig:fitPWA3}, showing overall a  good agreement between the data and fit model.

\begin{table}[h]
\caption{Dalitz fit results for magnitudes, phases and fit fractions (\%) of the spin-1 and spin-2 components, and the S-wave fit fraction. The uncertainties quoted are, in order,  statistical, experimental systematics, and model systematics. \label{tab:finalPWA3}}
\centering
\resizebox{16 cm}{!}{%
\begin{tabular}{lccc}
Component &  $\phantom{(}  $ Magnitude $ \phantom{)\times 10^{-2}}   $&Phase [$^\circ$]  & Fit fraction [\%]  \\\hline
$\rho(770)^0\pip$    & $\phantom{(} 1$ [fixed]       $ \phantom{)\times 10^{-2}}   $                         & $0 $ [fixed]                                & $    26.0 \phantom{22}   \pm    0.3   \phantom{22}   \pm    1.6 \phantom{22}     \pm    0.3 \phantom{22}    $\\
$\omega(782)\pip$    & $(1.68 \pm 0.06 \pm  0.15  \pm   0.02)\times 10^{-2}$  & $-103.3 \pm 2.1   \pm   2.6   \pm   0.4 $  & $\phantom{2}0.103   \pm    0.008   \pm    0.014   \pm    0.002  $\\
$\rho(1450)^0\pip$   & $\phantom{(} 2.66   \pm 0.07   \pm   0.24   \pm    0.22\phantom{)\times 10^{-2}} $  & $\phantom{-0}47.0   \pm 1.5   \pm   5.5   \pm    4.1 $  & $    \phantom{1}5.4  \phantom{22}    \pm    0.4   \phantom{22}   \pm    1.3    \phantom{22}  \pm    0.8   \phantom{22} $\\
$\rho(1700)^0\pip$   & $\phantom{(}7.41   \pm 0.18   \pm   0.47   \pm    0.71 \phantom{)\times 10^{-2}} $  & $-\phantom{0}65.7  \pm 1.5   \pm   3.8   \pm    4.6 $  & $    \phantom{1}5.7   \phantom{22}   \pm    0.5   \phantom{22}   \pm    1.0  \phantom{22}    \pm    1.0\phantom{22}    $\\
$f_2(1270)\pip$      & $\phantom{(}2.16   \pm 0.02   \pm   0.10   \pm    0.02 \phantom{)\times 10^{-2}} $  & $-100.9 \pm 0.7   \pm   2.0   \pm    0.4 $  & $    13.8  \phantom{22}   \pm    0.2  \phantom{22}    \pm    0.4    \phantom{22}  \pm    0.2  \phantom{22}  $\\
S-wave          &                                                          &                                    & $    61.8   \phantom{22}  \pm    0.5    \phantom{22}  \pm    0.6  \phantom{22}    \pm    0.5  \phantom{22}  $\\ \hline
  $\sum_i {\rm FF}_i$ & & & 112.8  \\
  $\chi^2$/ndof (range) & [1.47 - 1.78]&   & $-2\log{\mathcal{L}}$ = 805622 \\
\bottomrule
\end{tabular}
}
\end{table}

\begin{table}[h]
    \caption{Dalitz fit results for the interference fit fractions (\%) (statistical uncertainties only).} 
    \centering
    \resizebox{13 cm}{!}{%
    \begin{tabular}{c |rrrrr}
                  & {$\omega(782)\pip$} &  {$\rho(1450)^0\pip$} & {$\rho(1700)^0\pip$}          & {$f_2(1270)\pip$}           &  { S-wave}\\\hline
$\rho(770)^0\pip$ &  $-0.24\pm 0.06$  & $5.1\pm 0.3\phantom{1}$    & $-5.8\pm 0.4\phantom{1}$   & $-0.3\pm 0.1\phantom{1}$    & $1.8\pm 0.4\phantom{1}$ \\
$\omega(782)\pip$ &            & $0.05\pm 0.01$   & $0.05\pm 0.01$   & $0.046\pm 0.04$   & $-0.04\pm 0.01$ \\
$\rho(1450)^0\pip$&            &           & $-4.0\pm 0.5\phantom{1}$   & $1.1\pm 0.1\phantom{1}$     & $1.7\pm 0.2\phantom{1}$ \\
$\rho(1700)^0\pip$&            &           &           & $-0.8\pm 0.1\phantom{1}$    & $-3.4\pm 0.5\phantom{1}$ \\
$f_2(1270)\pip$   &            &           &           &            & $-1.6\pm 0.1\phantom{1}$\\
        \hline
    \end{tabular}
    }
    \label{tab:intFF}
\end{table}

\begin{figure}[h] 
 \centering
\includegraphics[width= 0.495\textwidth ]{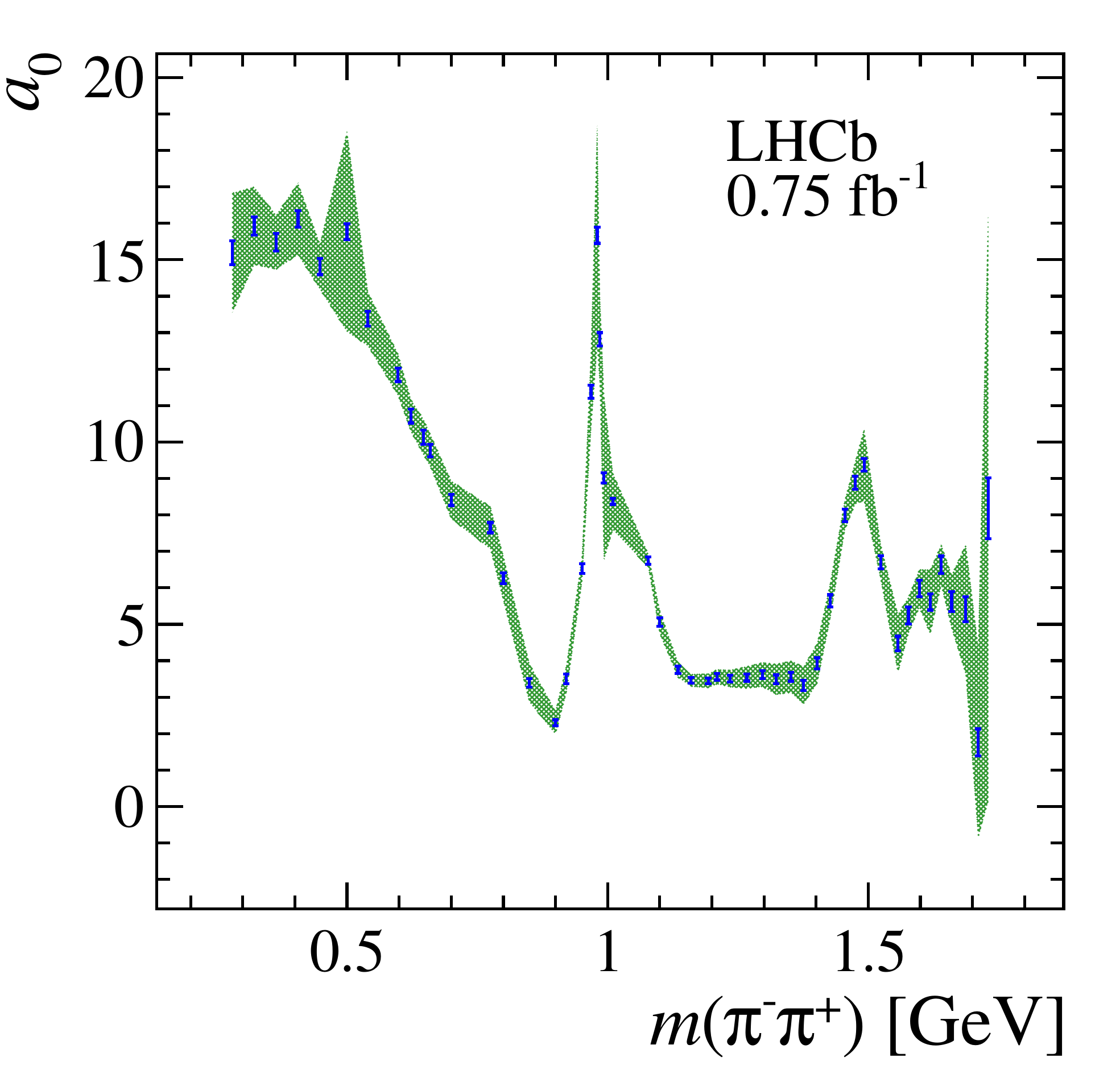}
\includegraphics[width= 0.495\textwidth]{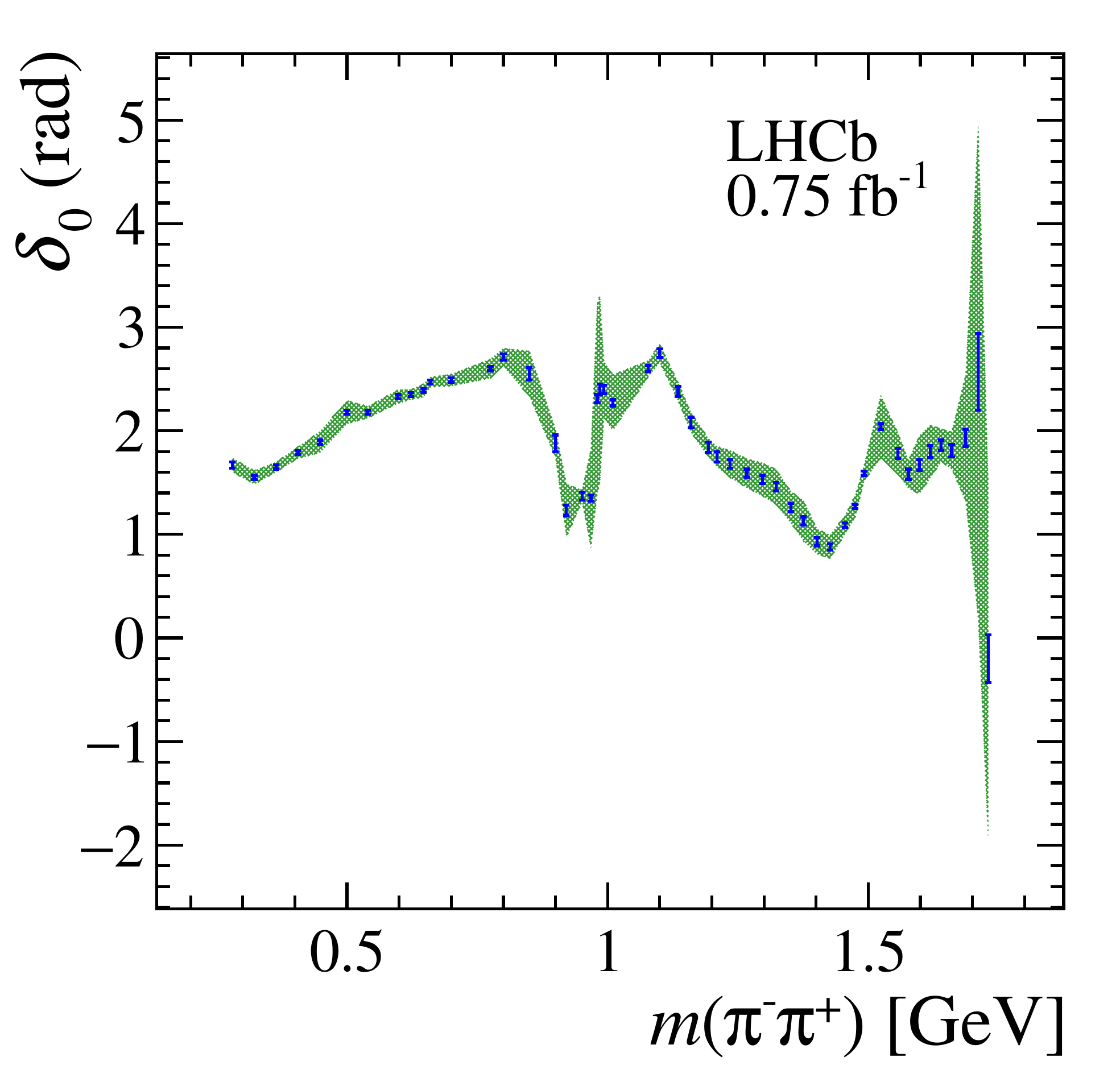}
\caption{Fitted (left) magnitude and (right) phase of the \pim\pip S-wave amplitude with statistical uncertainties as the blue bars and the total uncertainties (combined statistical, experimental and model systematics) as the green bands.}
\label{fig:S_allsyst}
\end{figure}

The S-wave component is found to be the dominant contribution, with a fraction of almost 62\%, in agreement with previous observations based on the isobar model. The $\rho(770)^0\pip$ channel is the second greatest contribution ($\sim 26\%)$, followed by the $f_2(1270)\pip$ mode ($\sim 14\%)$. The two heavy $\rho$ states contribute at the 5\% level, with a destructive interference pattern evidenced by an interference fit fraction of the same order. 

As anticipated by visual inspection, there is a small but clear contribution from the resonant state $\omega(782)$, with an interference pattern with $\rho(770)^0$, which can also be seen in the $s_{\pim\pip}$ projection in Fig.~\ref{fig:fitPWA3}. 
This contribution  was not observed in previous analyses due to their limited datasets, thus this effect is observed for the first time in this analysis. 
The  alternative parameterisation of the  $\rho-\omega$ mixing lineshape shown in Eq.~\ref{rhomix} was tested,  with $|B|$ and $\phi_B$ as free parameters. The outcome of the fit represents essentially the same solution: a difference in $-2\log{\cal L}$ of only $+1$ unit, no significant differences in all the other fit parameters, and the resulting  $\rho-\omega$ amplitude being almost indistinguishable to that of the sum of $\rho(770)^0$ and $\omega(782)$ isobars. The values found for $\rho-\omega$ mixing parameters are $|B| = 0.52 \pm 0.02\pm 0.05\pm 0.01$ and \mbox{$\phi_B= (158.8 \pm 2.1\pm 2.6   \pm   0.4)^\circ$}, where the quoted uncertainties are, in order, statistical, experimental systematics and model systematic uncertainties.

While the contribution of the $\rho(1450)^0$ resonance has been reported previously in the \Dppp decay~\cite{e791Dp3pi}, this state alone is not enough to describe the $\pim\pip$ P-wave amplitude at high mass. The inclusion of the $\rho(1700)^0$ resonance in the model improves the fit significantly, resulting in a difference in $-2\log{\cal L}$ of $-488$ units compared to that of the model without it. Its contribution is robust, with its stability in the fit being tested through many fit variations (as part of the systematic studies discussed in Sec.~\ref{sec:systematics}). The  $\rho(1700)^0$ is a broad state (nominal width of $250 \pm 100 \mev$ \cite{PDG2020}), and its inclusion affects the whole Dalitz plot, with an interesting interference pattern with the other vector states,  in particular with the $\rho(1450)^0$ resonance. 
The results presented here provide input for the debate of  $\rho(1450)^0-\rho(1700)^0$ interference,  supporting the need for these two overlapping states to exist (see for instance the entry for $\rho(1700)^0$ under ``Particle listings" in \cite{PDG2020}).
Note that if these states are excitations of the $\rho^0(770)$, their production should be also favoured in the $D^+$ decay.

The S-wave amplitude obtained from the final-model fit is shown in Fig.~\ref{fig:S_allsyst}. 
The values of the magnitude and phases for each knot are shown in Table~\ref{tab:mi_final_pwa_3}. The magnitude $a_0(m_{\pim\pip})$ is large close to threshold and decreases until $ \sim 0.9 \gev$, with a steady phase increase, as expected from the dominant $f_0(500)$ contribution reported in previous analyses, and consistent with a $d\bar d$ source of the \Dppp weak decay. Starting at $m_{\pim\pip}\sim 0.9\gev$ the $f_0(980)$ signature is observed both as a sharp increase of the magnitude and a rapid variation (decrease and increase) in the phase, enhanced possibly by the opening of the $K\Kb$ channel. Starting at about $m_{\pim\pip} \sim 1.4\gev$ and peaking near $m_{\pim\pip} \sim 1.5\gev$,  another structure in the amplitude with a corresponding phase movement is observed, indicating the presence of at least one more scalar resonance, possibly $f_0(1370)$ or $f_0(1500)$, or a combination of the two. 

\section{Systematic uncertainties}
\label{sec:systematics}

The systematic uncertainties on the fit parameters are divided into two categories: first those coming from the impact  of experimental aspects; and second, referred to as model systematics, those corresponding to the uncertainties in resonance lineshape parameters such as masses, widths and form factors. 

In the first category, effects due to the efficiency modelling, background contributions and intrinsic fit biases are considered. For the efficiency map, uncertainties arise from the finite size of the simulation samples, the effect of the binning scheme of the efficiency histogram prior to the 2D spline smoothing, and the procedure for obtaining PID efficiencies from the calibration samples. The first effect is estimated by generating a set of alternative efficiency histograms where the bin contents are varied according to a Poisson distribution, and performing the Dalitz fit with each alternative map. The root mean square (rms) of the distribution of each of the fit parameters is assigned as the corresponding systematic uncertainty. The systematic uncertainty due to the binning scheme of the efficiency map is assessed by performing the fits with smoothed efficiency maps obtained from varying the binning grid of the efficiency histogram from $15\times 15$ (default) to $12\times 12$ and $20\times 20$. The largest variation in each fit parameter is assigned as the systematic uncertainty. The dominant uncertainties on the PID efficiency are due to the finite size of the calibration samples; the effect is assessed by varying the efficiencies from the calibration tables according to a Gaussian distribution centered at the nominal values and with widths equal to the statistical uncertainties, and obtaining new smoothed efficiency maps to refit the Dalitz plot and assess the impact on the fit parameters. 

The effect of the signal-to-background ratio is estimated by varying the signal purity obtained from the invariant-mass fit within one standard deviation, repeating the Dalitz fit and assigning the difference in each parameter as the systematic uncertainty. The impact of the background model is addressed by determining it from either just the lower or just the higher mass sideband, and assigning the largest deviation in each fit parameter as the corresponding systematic uncertainty. The S-wave interval definition (knots in $m_{\pim\pip}$) is varied both in the number of knots (from 45 to 55) and position of the knots,  by using an adaptive binning where the division is made to equalise the event population. The resulting S-wave amplitude in each case after the fit is used to calculate the values of magnitude and phase in the original knot scheme and the rms in each  fit parameter (including those from spin-1 and spin-2 states) is assigned as systematic uncertainty.  

Finally, uncertainties due to biases from the fitting algorithm, including the ability to reproduce a given (continuous) S-wave amplitude,  are assessed by generating a large number of pseudoexperiments given by the full PDF (signal and background). The PDF for the signal model has its parameters set to those obtained from a QMIPWA fit with an alternative 50-knot choice using    the  adaptive binning. The pseudoexperiments are fitted with the baseline 50-knot model, and the rms of the distribution of the difference in each parameter is assigned as the systematic uncertainty. Overall, the largest systematic uncertainties come from the finite size of the simulation samples, the efficiency map binning, and the fit bias. The systematic uncertainties are all added in quadrature and comprise the second uncertainty in Tables~\ref{tab:finalPWA3} and \ref{tab:mi_final_pwa_3}.
Other studies are performed as cross-checks and lead to variations within statistical uncertainties: the use of the efficiency map without correcting for mismatches between data and simulation; splitting the sample in terms of magnet orientation during data taking; and studying the effect of mass resolution around the region of the narrow $\omega(782)$ state.

In the second category, isobar model systematics, two studies are performed. First, the mass and the width of all the spin-1 and spin-2 states are varied within their quoted uncertainties \cite{PDG2020}, and the rms of the difference in each fit parameter is assigned as a systematic uncertainty. The second study addresses the impact of changing the effective radii of both the resonances and the $D^+$ meson used in the Blatt-Weisskopf barrier factors, to 1.0 and 2.0 GeV$^{-1}$  and to 4.0 and 6.0 GeV$^{-1}$, respectively, and the largest variation for each parameter is assigned as systematic uncertainty. These two effects are added in quadrature and comprise the third uncertainties presented in Tables~\ref{tab:finalPWA3} and \ref{tab:mi_final_pwa_3}.

\begin{table}
\caption{ Fitted magnitude and phase of the S-wave amplitude at each \pim\pip mass knot, relative to the  $\rho(770)^0\pip$ channel. The uncertainties quoted are, in order,  statistical, experimental systematics and model systematics. }
\centering
\resizebox{12 cm}{!}{%
\begin{tabular}{cccc}
knot & $m_{\pim\pip}$ [\gev] &  Magnitude &  Phase [rad]  \\
\toprule                                         
1     &    0.280    &$    15.2   \pm    0.3  \pm    1.5    \pm    0.5    $&$    \phantom{-}1.67    \pm    0.03    \pm    0.06    \pm    0.02    $\\
2     &    0.322    &$    15.9   \pm    0.3  \pm    0.9    \pm    0.4    $&$    \phantom{-}1.55    \pm    0.02    \pm    0.06    \pm    0.02    $\\
3     &    0.364    &$    15.5   \pm    0.2  \pm    0.6    \pm    0.4    $&$    \phantom{-}1.65    \pm    0.02    \pm    0.04    \pm    0.02    $\\
4     &    0.406    &$    16.1   \pm    0.2  \pm    0.9    \pm    0.4    $&$    \phantom{-}1.79    \pm    0.02    \pm    0.05    \pm    0.02    $\\
5     &    0.448    &$    14.8   \pm    0.2  \pm    0.5    \pm    0.3    $&$    \phantom{-}1.89    \pm    0.02    \pm    0.10    \pm    0.02    $\\
6     &    0.500    &$    15.8   \pm    0.2  \pm    2.7    \pm    0.3    $&$    \phantom{-}2.18    \pm    0.02    \pm    0.11    \pm    0.02    $\\
7     &    0.540    &$    13.4   \pm    0.2  \pm    0.7    \pm    0.3    $&$    \phantom{-}2.18    \pm    0.02    \pm    0.05    \pm    0.02    $\\
8     &    0.598    &$    11.8   \pm    0.2  \pm    0.5    \pm    0.3    $&$    \phantom{-}2.33    \pm    0.02    \pm    0.06    \pm    0.02    $\\
9     &    0.623    &$    10.7   \pm    0.2  \pm    0.3    \pm    0.2    $&$    \phantom{-}2.35    \pm    0.02    \pm    0.04    \pm    0.02    $\\
10    &    0.647    &$    10.1   \pm    0.2  \pm    0.4    \pm    0.2    $&$    \phantom{-}2.39    \pm    0.02    \pm    0.06    \pm    0.02    $\\
11    &    0.660    &$    \phantom{1}9.8    \pm    0.2  \pm    0.4    \pm    0.2    $&$    \phantom{-}2.47    \pm    0.02    \pm    0.04    \pm    0.02    $\\
12    &    0.700    &$    \phantom{1}8.4    \pm    0.2  \pm    0.5    \pm    0.2    $&$    \phantom{-}2.49    \pm    0.02    \pm    0.05    \pm    0.02    $\\
13    &    0.775    &$    \phantom{1}7.7    \pm    0.1  \pm    0.6    \pm    0.1    $&$    \phantom{-}2.60    \pm    0.02    \pm    0.09    \pm    0.01    $\\
14    &    0.800    &$    \phantom{1}6.3    \pm    0.2  \pm    0.6    \pm    0.1    $&$    \phantom{-}2.71    \pm    0.03    \pm    0.08    \pm    0.02    $\\
15    &    0.850    &$    \phantom{1}3.4    \pm    0.1  \pm    0.5    \pm    0.1    $&$    \phantom{-}2.55    \pm    0.06    \pm    0.21    \pm    0.03    $\\
16    &    0.900    &$    \phantom{1}2.3    \pm    0.1  \pm    0.3    \pm    0.1    $&$    \phantom{-}1.88    \pm    0.08    \pm    0.10    \pm    0.05    $\\
17    &    0.921    &$    \phantom{1}3.5    \pm    0.1  \pm    0.3    \pm    0.1    $&$    \phantom{-}1.23    \pm    0.05    \pm    0.25    \pm    0.03    $\\
18    &    0.951    &$    \phantom{1}6.5    \pm    0.1  \pm    0.3    \pm    0.1    $&$    \phantom{-}1.37    \pm    0.04    \pm    0.04    \pm    0.02    $\\
19    &    0.968    &$    11.4   \pm    0.2  \pm    1.0    \pm    0.1    $&$    \phantom{-}1.35    \pm    0.03    \pm    0.48    \pm    0.01    $\\
20    &    0.980    &$    15.7   \pm    0.2  \pm    3.0    \pm    0.1    $&$    \phantom{-}2.31    \pm    0.04    \pm    0.92    \pm    0.01    $\\
21    &    0.985    &$    12.8   \pm    0.2  \pm    1.2    \pm    0.1    $&$    \phantom{-}2.40    \pm    0.05    \pm    0.91    \pm    0.01    $\\
22    &    0.993    &$    \phantom{1}9.0    \pm    0.1  \pm    2.6    \pm    0.0    $&$    \phantom{-}2.40    \pm    0.04    \pm    0.27    \pm    0.01    $\\
23    &    1.010    &$    \phantom{1}8.4    \pm    0.1  \pm    0.8    \pm    0.1    $&$    \phantom{-}2.27    \pm    0.03    \pm    0.27    \pm    0.01    $\\
24    &    1.078    &$    \phantom{1}6.8    \pm    0.1  \pm    0.1    \pm    0.1    $&$    \phantom{-}2.60    \pm    0.03    \pm    0.07    \pm    0.02    $\\
25    &    1.100    &$    \phantom{1}5.1    \pm    0.1  \pm    0.3    \pm    0.1    $&$    \phantom{-}2.75    \pm    0.04    \pm    0.08    \pm    0.02    $\\
26    &    1.135    &$    \phantom{1}3.8    \pm    0.1  \pm    0.2    \pm    0.1    $&$    \phantom{-}2.38    \pm    0.05    \pm    0.07    \pm    0.03    $\\
27    &    1.160    &$    \phantom{1}3.5    \pm    0.1  \pm    0.1    \pm    0.1    $&$    \phantom{-}2.08    \pm    0.05    \pm    0.08    \pm    0.03    $\\
28    &    1.193    &$    \phantom{1}3.5    \pm    0.1  \pm    0.1    \pm    0.1    $&$    \phantom{-}1.84    \pm    0.05    \pm    0.07    \pm    0.03    $\\
29    &    1.210    &$    \phantom{1}3.6    \pm    0.1  \pm    0.1    \pm    0.1    $&$    \phantom{-}1.75    \pm    0.05    \pm    0.08    \pm    0.03    $\\
30    &    1.235    &$    \phantom{1}3.5    \pm    0.1  \pm    0.2    \pm    0.1    $&$    \phantom{-}1.68    \pm    0.04    \pm    0.12    \pm    0.04    $\\
31    &    1.267    &$    \phantom{1}3.5    \pm    0.1  \pm    0.3    \pm    0.1    $&$    \phantom{-}1.59    \pm    0.04    \pm    0.13    \pm    0.04    $\\
32    &    1.297    &$    \phantom{1}3.6    \pm    0.1  \pm    0.3    \pm    0.2    $&$    \phantom{-}1.53    \pm    0.04    \pm    0.15    \pm    0.04    $\\
33    &    1.323    &$    \phantom{1}3.5    \pm    0.1  \pm    0.4    \pm    0.2    $&$    \phantom{-}1.46    \pm    0.04    \pm    0.16    \pm    0.05    $\\
34    &    1.352    &$    \phantom{1}3.6    \pm    0.1  \pm    0.4    \pm    0.2    $&$    \phantom{-}1.26    \pm    0.04    \pm    0.14    \pm    0.05    $\\
35    &    1.376    &$    \phantom{1}3.3    \pm    0.1  \pm    0.5    \pm    0.2    $&$    \phantom{-}1.13    \pm    0.04    \pm    0.18    \pm    0.07    $\\
36    &    1.402    &$    \phantom{1}3.9    \pm    0.2  \pm    0.5    \pm    0.2    $&$    \phantom{-}0.93    \pm    0.04    \pm    0.10    \pm    0.06    $\\
37    &    1.427    &$    \phantom{1}5.6    \pm    0.2  \pm    0.4    \pm    0.2    $&$    \phantom{-}0.88    \pm    0.03    \pm    0.10    \pm    0.05    $\\
38    &    1.455    &$    \phantom{1}8.0    \pm    0.2  \pm    0.3    \pm    0.2    $&$    \phantom{-}1.09    \pm    0.02    \pm    0.08    \pm    0.04    $\\
39    &    1.475    &$    \phantom{1}8.9    \pm    0.2  \pm    0.5    \pm    0.2    $&$    \phantom{-}1.27    \pm    0.02    \pm    0.10    \pm    0.04    $\\
40    &    1.492    &$    \phantom{1}9.4    \pm    0.2  \pm    0.9    \pm    0.3    $&$    \phantom{-}1.59    \pm    0.02    \pm    0.07    \pm    0.03    $\\
41    &    1.524    &$    \phantom{1}6.7    \pm    0.2  \pm    0.3    \pm    0.4    $&$    \phantom{-}2.04    \pm    0.03    \pm    0.30    \pm    0.04    $\\
42    &    1.557    &$    \phantom{1}4.5    \pm    0.2  \pm    0.7    \pm    0.3    $&$    \phantom{-}1.78    \pm    0.05    \pm    0.18    \pm    0.08    $\\
43    &    1.577    &$    \phantom{1}5.2    \pm    0.2  \pm    0.4    \pm    0.3    $&$    \phantom{-}1.58    \pm    0.05    \pm    0.10    \pm    0.08    $\\
44    &    1.598    &$    \phantom{1}6.0    \pm    0.2  \pm    0.4    \pm    0.3    $&$    \phantom{-}1.67    \pm    0.05    \pm    0.27    \pm    0.07    $\\
45    &    1.619    &$    \phantom{1}5.6    \pm    0.2  \pm    0.8    \pm    0.4    $&$    \phantom{-}1.80    \pm    0.06    \pm    0.24    \pm    0.07    $\\
46    &    1.640    &$    \phantom{1}6.6    \pm    0.2  \pm    0.3    \pm    0.4    $&$    \phantom{-}1.86    \pm    0.05    \pm    0.14    \pm    0.06    $\\
47    &    1.660    &$    \phantom{1}5.6    \pm    0.3  \pm    0.5    \pm    0.4    $&$    \phantom{-}1.81    \pm    0.06    \pm    0.15    \pm    0.08    $\\
48    &    1.687    &$    \phantom{1}5.4    \pm    0.3  \pm    1.7    \pm    0.5    $&$    \phantom{-}1.93    \pm    0.08    \pm    0.60    \pm    0.10    $\\
49    &    1.711    &$    \phantom{1}1.8    \pm    0.4  \pm    2.5    \pm    0.6    $&$    \phantom{-}2.57    \pm    0.37    \pm    2.33    \pm    0.34    $\\
50    &    1.730    &$    \phantom{1}8.2    \pm    0.8  \pm    8.0    \pm    0.7    $&$    -0.20    \pm    0.23    \pm    1.71    \pm    0.04    $\\
\bottomrule
\end{tabular}
}
\label{tab:mi_final_pwa_3}
\end{table}

\section{Summary and conclusions}
\label{sec:conclusion}

This paper presents a Dalitz plot analysis of the \Dppp decay. Using a sample containing more than  six hundred thousand candidates, with a purity of 95\%, the resonant structure of the decay is studied using the QMIPWA method, where the magnitude and the phase of the S-wave amplitude is obtained as a function of $m_{\pim\pip}$, while the spin-1 and spin-2 contributions are included with an isobar model.  This approach is motivated by the presence of broad and overlapping light scalar resonances below 2\gev, with poorly known masses and widths.

The result of the Dalitz plot fit shows that the decay is dominated by the $\pim\pip$ S-wave component corresponding to nearly 62\% of the \Dppp decay rate,  which is consistent with results from previous analyses. The P-wave amplitude is the second largest component, led by the $\rho(770)^0$  contribution at the level of 26\% but including also the $\omega(782)$ resonance and the high-mass states $\rho(1450)^0$ and $\rho(1700)^0$. The D-wave amplitude, consisting only of the $f_2(1270)$ state, accounts for about 14\% of the \Dppp decay rate.

The contribution of $D^+\to \omega(782)\pi^+$ in the \Dppp decay is observed for the first time, with a fit fraction of  $(0.103\pm 0.016)\%$. 
The $\omega(782)\to \pim\pip$ decay is isospin violating and has been observed in association with the $\rho(770)\to \pim\pip$ in a few processes with very different interference patterns \cite{KLOE2003,CMD-2_eepipi_2007,CMD2-2002,CDF-X3872-rho-omega,Belle-X3872-rho-omega,BESIII_etap_gammapipi_2018,LHCb-PAPER-2019-017,LHCb-PAPER-2021-045}.

Interesting structures are observed in the S-wave amplitude shown in Fig.~\ref{fig:S_allsyst}. The broad structure near threshold is associated with the $f_0(500)$ resonance in model-dependent analyses, the $f_0(980)$ state is clearly visible with an asymmetric peak lineshape, and a further peak with corresponding phase variation is observed around $1.5\gev$  indicating the presence of a further high-mass scalar state. 

 While this analysis  is  based on the concept of a well-isolated companion pion, that is, a 2+1 approximation, the capacity of the QMIPWA approach to absorb some three-body final-state interaction effects should be considered in the interpretation of the resulting (2-body) \pim\pip S-wave amplitude obtained from the fit.

 This is the first time  that the $\pim\pip$ S-wave amplitude is extracted through a quasi-model-independent approach for the \Dppp decay, from threshold up to $1.7\gev$. Together with the companion analysis of the \Dsppp channel\cite{LHCb-PAPER-2022-030} this provide an important input to phenomenological studies.

%% file: acknowledgements.tex
\section*{Acknowledgements}
%
%
\noindent We express our gratitude to our colleagues in the CERN
accelerator departments for the excellent performance of the LHC. We
thank the technical and administrative staff at the LHCb
institutes.
We acknowledge support from CERN and from the national agencies:
CAPES, CNPq, FAPERJ and FINEP (Brazil); 
MOST and NSFC (China); 
CNRS/IN2P3 (France); 
BMBF, DFG and MPG (Germany); 
INFN (Italy); 
NWO (Netherlands); 
MNiSW and NCN (Poland); 
MEN/IFA (Romania); 
MICINN (Spain); 
SNSF and SER (Switzerland); 
NASU (Ukraine); 
STFC (United Kingdom); 
DOE NP and NSF (USA).
We acknowledge the computing resources that are provided by CERN, IN2P3
(France), KIT and DESY (Germany), INFN (Italy), SURF (Netherlands),
PIC (Spain), GridPP (United Kingdom), 
CSCS (Switzerland), IFIN-HH (Romania), CBPF (Brazil),
Polish WLCG  (Poland) and NERSC (USA).
We are indebted to the communities behind the multiple open-source
software packages on which we depend.
Individual groups or members have received support from
ARC and ARDC (Australia);
Minciencias (Colombia);
AvH Foundation (Germany);
EPLANET, Marie Sk\l{}odowska-Curie Actions and ERC (European Union);
A*MIDEX, ANR, IPhU and Labex P2IO, and R\'{e}gion Auvergne-Rh\^{o}ne-Alpes (France);
Key Research Program of Frontier Sciences of CAS, CAS PIFI, CAS CCEPP, 
Fundamental Research Funds for the Central Universities, 
and Sci. \& Tech. Program of Guangzhou (China);
GVA, XuntaGal, GENCAT and Prog.~Atracci\'on Talento, CM (Spain);
SRC (Sweden);
the Leverhulme Trust, the Royal Society
 and UKRI (United Kingdom).

%% file: Authorship_LHCb-PAPER-2022-016.tex
\centerline
{\large\bf LHCb collaboration}
\begin
{flushleft}
\small
R.~Aaij$^{32}$\lhcborcid{0000-0003-0533-1952},
A.S.W.~Abdelmotteleb$^{50}$\lhcborcid{0000-0001-7905-0542},
C.~Abellan~Beteta$^{44}$,
F.~Abudin{\'e}n$^{50}$\lhcborcid{0000-0002-6737-3528},
T.~Ackernley$^{54}$\lhcborcid{0000-0002-5951-3498},
B.~Adeva$^{40}$\lhcborcid{0000-0001-9756-3712},
M.~Adinolfi$^{48}$\lhcborcid{0000-0002-1326-1264},
H.~Afsharnia$^{9}$,
C.~Agapopoulou$^{13}$\lhcborcid{0000-0002-2368-0147},
C.A.~Aidala$^{76}$\lhcborcid{0000-0001-9540-4988}
% ...

{\footnotesize \it

% ...
$^{9}$Universit{\'e} Clermont Auvergne, CNRS/IN2P3, LPC, Clermont-Ferrand, France\\
% ...
$^{13}$LPNHE, Sorbonne Universit{\'e}, Paris Diderot Sorbonne Paris Cit{\'e}, CNRS/IN2P3, Paris, France\\
% ...
$^{32}$Nikhef National Institute for Subatomic Physics, Amsterdam, Netherlands\\
% ...
$^{40}$Instituto Galego de F{\'\i}sica de Altas Enerx{\'\i}as (IGFAE), Universidade de Santiago de Compostela, Santiago de Compostela, Spain\\
% ...
$^{44}$Physik-Institut, Universit{\"a}t Z{\"u}rich, Z{\"u}rich, Switzerland\\
% ...
$^{48}$H.H. Wills Physics Laboratory, University of Bristol, Bristol, United Kingdom\\
% ...
$^{50}$Department of Physics, University of Warwick, Coventry, United Kingdom\\
% ...
$^{54}$Oliver Lodge Laboratory, University of Liverpool, Liverpool, United Kingdom\\
% ...
$^{62}$Syracuse University, Syracuse, NY, United States\\
% ...
$^{76}$University of Michigan, Ann Arbor, MI, United States, associated to $^{62}$\\
\bigskip
% ...
}
\end{flushleft}